\documentclass[useAMS,usenatbib,aas_macros]{mn2e} 
\usepackage{graphicx}
\usepackage {epsfig,times,aas_macros}
\usepackage{amssymb}

\newcommand{\se}[1]{\S\ref{sec:#1}}
\newcommand{\fig}[1]{Fig.~\ref{fig:#1}}

\newcommand{\Fig}[1]{Figure~\ref{fig:#1}}

\newcommand{\be}{\begin{equation}}
\newcommand{\ee}{\end{equation}}
\newcommand{\bea}{\begin{eqnarray}}
\newcommand{\eea}{\end{eqnarray}}

\newcommand{\msun}{{\rm M}_\odot}

\newcommand{\ifm}[1]{\relax\ifmmode#1\else$\mathsurround=0pt #1$\fi}
\newcommand{\kms}{\ifmmode\,{\rm km}\,{\rm s}^{-1}\else km$\,$s$^{-1}$\fi}
\newcommand{\hmpc}{\,\ifm{h^{-1}}{\rm Mpc}}

\newcommand{\Mpc}{\,{\rm Mpc}}
\newcommand{\kpc}{\,{\rm kpc}}
\newcommand{\pc}{\,{\rm pc}}

\newcommand{\Myr}{\,{\rm Myr}}

\newcommand{\ltsima}{$\; \buildrel < \over \sim \;$}
\newcommand{\lsim}{\lower.5ex\hbox{\ltsima}}
\newcommand{\gtsima}{$\; \buildrel > \over \sim \;$}
\newcommand{\gsim}{\lower.5ex\hbox{\gtsima}}

\def\la{\langle}
\def\ra{\rangle}

\def\omm{\Omega_{\rm m}}
\def\oml{\Omega_{\Lambda}}
\def\omb{\Omega_{\rm b}}

\def\syr{\, M_\odot\,{\rm yr}^{-1}\,{\rm rad}^{-2}}

\def\cmc{\,{\rm cm}^{-3}}

\def\Mv{M_{\rm v}}
\def\Rv{R_{\rm v}}
\def\Vv{V_{\rm v}}

\def\rs{r_{\rm s}}

\def\rd0{r_{{\rm d}0}}
\def\Vdi0{V_{{\rm d}0}}

\usepackage{color}
\newcommand{\adb}[1]{#1}

\begin{document}

\large

\title[Streams, pancakes and galaxy angular momentum]
{ 
Co-planar streams, pancakes, and angular-momentum exchange in high-z 
disc galaxies
}

\author[Danovich et al.]
{\parbox[t]{\textwidth}
{
Mark Danovich$^1$ \thanks{E-mail: mark.danovich@mail.huji.ac.il},
Avishai Dekel$^1$\thanks{E-mail: avishai.dekel@huji.ac.il},
Oliver Hahn$^2$,
Romain Teyssier$^{3,4}$ 
}
\\ \\ 
$^1$Racah Institute of Physics, The Hebrew University, Jerusalem 91904,
Israel\\
$^2$Kavli Institute for Particle Astrophysics and Cosmology, SLAC/Stanford
University, 2575 Sand Hill Road, Menlo Park, CA 94025, USA \\
$^3$CEA, IRFU, SAp, 91191 Gif-sur-Yvette, France\\
$^4$Institute for Theoretical Physics, University of Zurich, CH-8057 Zurich,
Switzerland
}
\date{}

\pagerange{\pageref{firstpage}--\pageref{lastpage}} \pubyear{0000}

\maketitle

\label{firstpage}

\begin{abstract}
We study the feeding of massive galaxies at high redshift through streams from 
the cosmic web using the Mare Nostrum hydro-cosmological simulation. Our 
statistical sample consists of 350 dark-matter haloes of $\simeq\!10^{12}\msun$
at $z\!=\!2.5$.  We find that $\sim\!70\%$ of the influx into the virial radius
$\Rv$ is in narrow streams covering $10\%$ of the virial shell. On average 64\%
of the stream influx is in one stream, and  95\% is in three dominant streams.
The streams that feed a massive halo tend to lie in a plane that extends from
half to a few $\Rv$, hereafter ``the stream plane" (SP). The streams are 
typically embedded in a thin sheet of low-entropy gas, a Zel'dovich pancake, 
which carries $\sim\!20\%$ of the influx into $\Rv$. The filaments-in-a-plane 
configuration about the massive haloes at the nodes of the cosmic web differs 
from the large-scale structure of the web where the filaments mark the 
intersections of slanted sheets.  The stream plane is only weakly aligned with 
the angular momentum (AM) near $\Rv$, consistent with the fact that typically 
80\% of the AM is carried by one dominant stream. The galactic disc plane shows
a weak tendency to be perpendicular to the large-scale SP, consistent with 
tidal-torque theory. Most interesting, the direction of the disc AM is only 
weakly correlated with the AM direction at $\Rv$. This indicates a significant 
AM exchange at the interphase between streams and disc in the greater 
environment of the disc inside an ``AM sphere of radius $\sim\!0.3\Rv$.  The 
required large torques are expected based on the perturbed morphology and 
kinematics within this interaction sphere.  This AM exchange may or may not 
require a major modification of the standard disc modeling based on AM 
conservation, depending on the extent to which the amplitude of the disc AM is 
affected, which is yet to be studied.
\end{abstract}


\section{Introduction}
\label{sec:intro}

\begin{figure*}
\begin{minipage}[b]{0.5\linewidth}
 \centering
\includegraphics[width=1.03\textwidth]{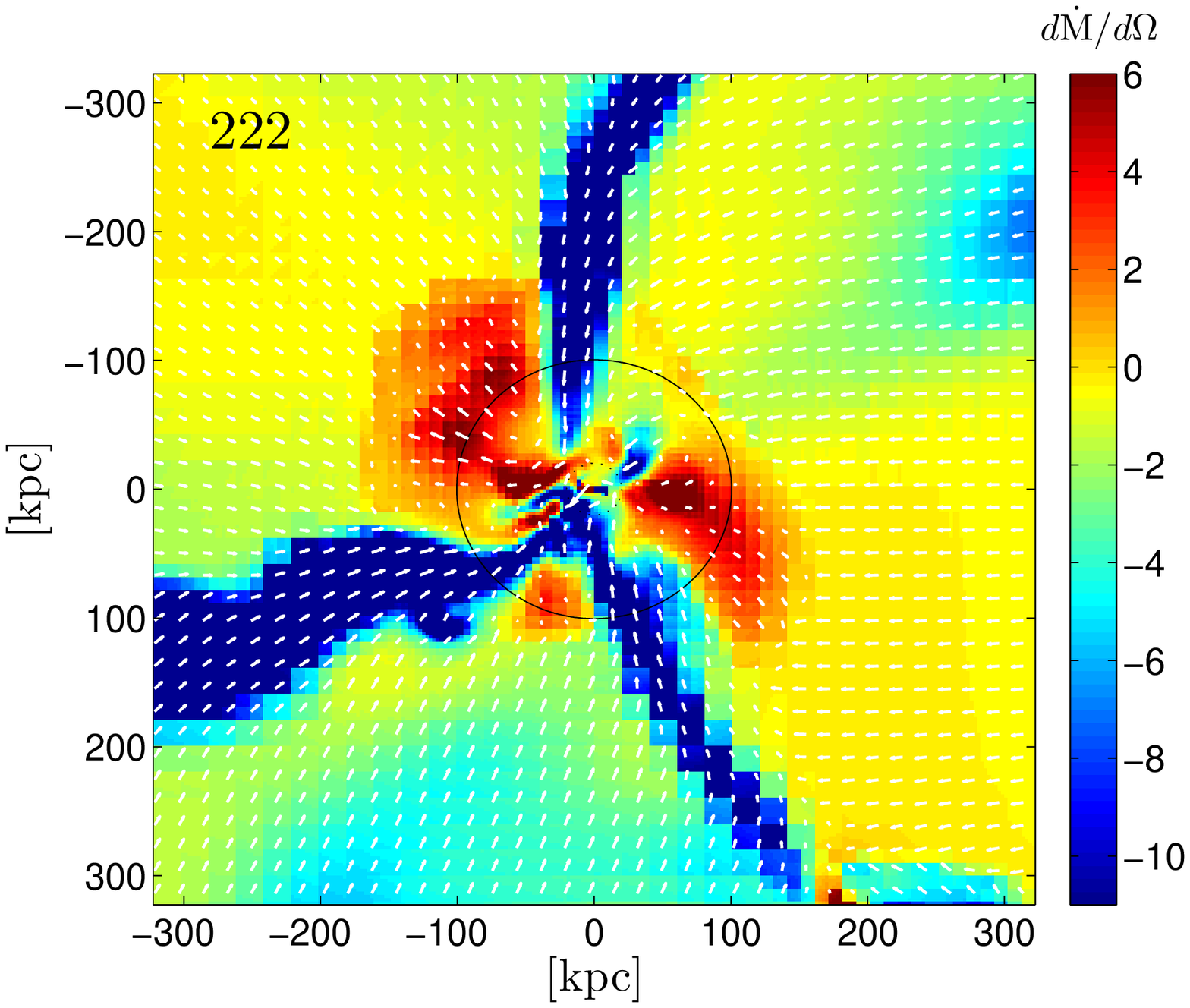}
\end{minipage}\hfill
\begin{minipage}[b]{0.47\linewidth}
 \centering
 \includegraphics[width=0.90\textwidth]{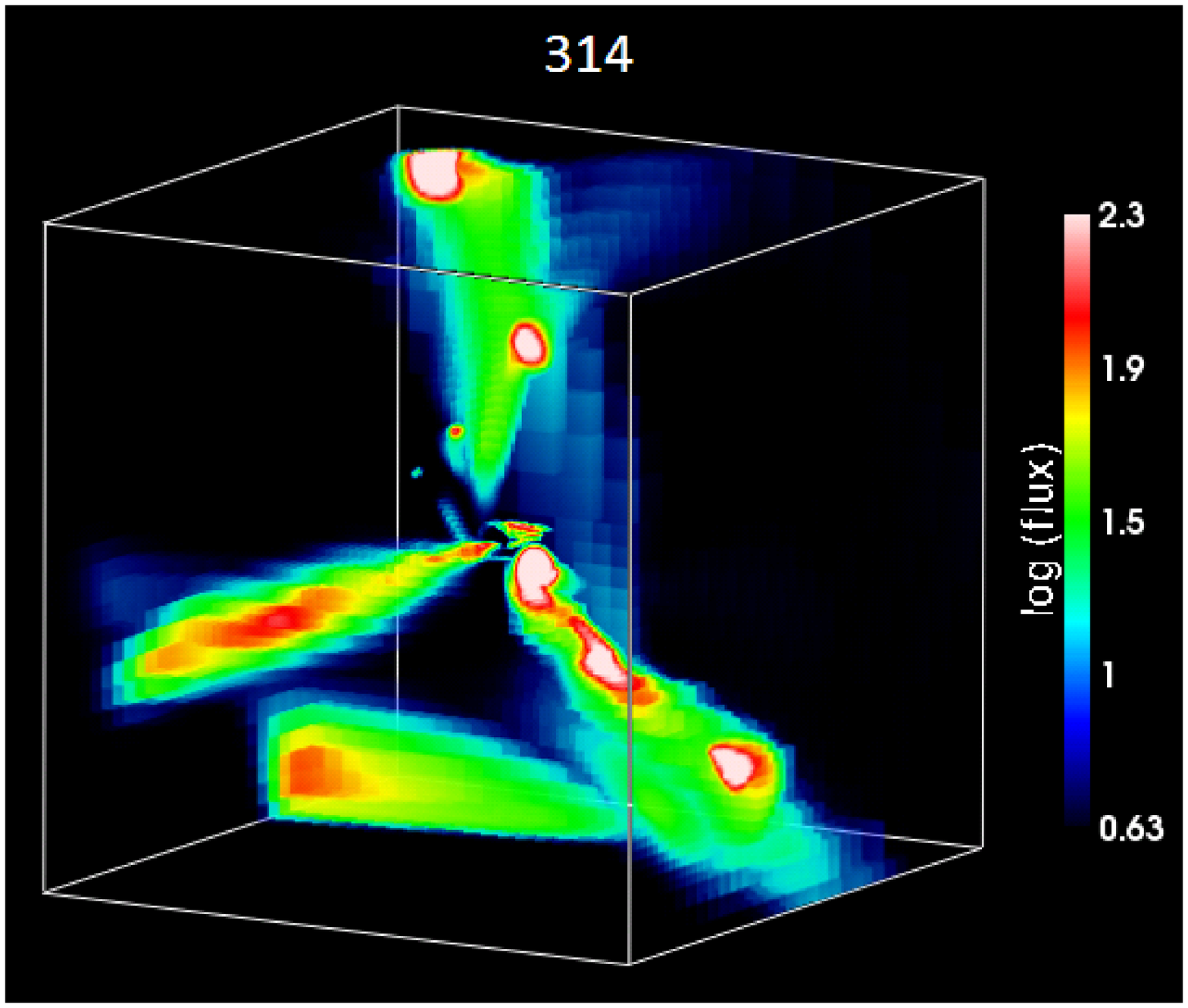}
\end{minipage}\hfill\\
\caption{
Cold gas streams. Shown is the radial flux density per solid angle 
\adb{(in $\syr$)}
into two 
\adb{typical} 
massive galaxies at $z=2.5$ from the Mare Nostrum cosmological simulation.
{\bf Left:} A 1-kpc thick slice through the centre of a halo,
showing three streams
that emerge at a distance more than 5 times the virial radius (black circle)
and penetrate to the inner halo.
{\bf Right:} a three-dimensional view 
\adb{of influx},
extending to twice the virial radius,
showing that the streams contain \adb{clumps} and a smoother gas component.
\adb{The clumps are merging galaxies containing gas, stars and dark matter,
\citep[see][]{dekel09,cdb10}.} 
}
\label{fig:proj}
\end{figure*}

\begin{figure}
\centering
\includegraphics[width=0.42\textwidth]{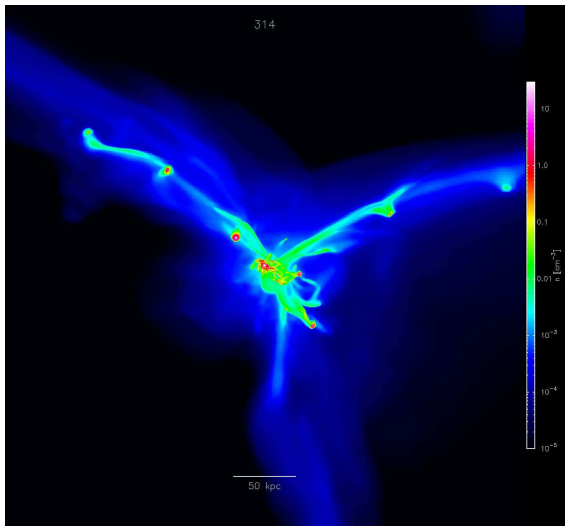}
\caption{
Gas density in the streams feeding the galaxy shown in the right panel of
\fig{proj} (314), 
in a 
\adb{RAMSES}
test run with a resolution higher by an order of magnitude.
\adb{The box
encompasses a sphere of twice the virial radius.}
The stream configuration remains 
\adb{qualitatively similar, with the inner structure of the
streams better resolved (courtesy of D.~Tweed).}
}
\label{fig:bird}
\end{figure}

The large-scale distribution of matter in the universe, dominated by the dark
matter, exhibits a cosmic web \citep{bond96,pogosyan98}. 
This has been envisioned based on the Zel'dovich
approximation \citep{zeldovich70} and the theory of Gaussian random fields
\citep{doroshkevich70,bbks}, reproduced in cosmological N-body simulations
\citep{shandarin-klypin, davis85, millenium, bolshoi}, 
and observed in redshift surveys
\citep{cfa,2df,sdss}. 
The honey-comb-like structure consists of big voids surrounded by flat 
low-overdensity sheets (walls, pancakes), that intersect in narrow, 
denser filaments, which further intersect in relatively compact, virialized, 
spheroidal dark-matter haloes (clusters) at the nodes of the web.   
According the quasi-linear Zel'dovich approximation \citep{zeldovich70},
each Lagrangian point is characterized by the eigenvectors and eigenvalues
of the deformation tensor, which consists of the second spatial partial
derivatives of the gravitational potential field. 
if we denote the eigenvalues $(\lambda_1,\lambda_2,\lambda_3)$, ordered
such that $\lambda_1\geq \lambda_2 \geq \lambda_3$,
a sheet would form due to a one-dimensional collapse where $\lambda_1$
dominates, which we can crudely approximate by the requirement that
the signature of the eigenvalues is $(+,-,-)$.
A filament would form after collapse in two directions,
where $\lambda_1$ and $\lambda_2$ are comparable and much larger than
$\lambda_3$, e.g., a signature $(+,+,-)$. 
A halo would form in a node, where the eigenvalues
are all comparable and positive, $(+,+,+)$.
The filaments thus form in the intersections of sheets, and the nodes are
intersections of filaments. In the quasi-linear regime, the matter flows
away from the voids into the sheets, within the sheets toward the filaments,
and along the filaments toward the nodes.

The filaments are the prominent features of the cosmic web,
both in terms of mass content and density contrast.
In the linear regime of a Gaussian random density fluctuation field, the
fraction of volume and mass  
associated with each of the four different signatures of the deformation tensor
is 42\% for the filaments and sheets, and 8\% for the haloes and voids 
\citep{doroshkevich70}.
In the non-linear regime, the statistics depend on scale,
as the structures have collapsed bottom-up to increasingly higher densities 
occupying smaller and smaller scales.
\citet{hahn07a} have demonstrated this scale dependence by showing how the 
volume fractions of the four signatures of the deformation tensor deviate from 
the Gaussian random field values as the smoothing scale is decreased.
Using adaptive filtering, \citet{aragon10} found that the mass fractions
associated with the four signatures of the Hessian tensor, 
consisting of the spatial partial derivatives of the density field,
are 28.1\% in haloes, 39.2\% in filaments, 5.5\% in sheets and 27.2\% 
in voids, while these structures occupy 0.4\%, 8.8\%, 4.9\% and 85.9\% of 
the volume, respectively. The corresponding mean overdensities with respect 
to the mean cosmological background are 73, 4.5, 1.1 and 0.3, respectively.
Using an excursion-set model for triaxial collapse, \citet{shen06} found
that at $z=0$, when filtering over scales corresponding to $10^{10}\,M_\odot$, 
typically 99\% of the mass has crossed the threshold for collapse along 
at least one axis, and thus resides in sheets (including filaments and haloes),
while 72\% is in filaments (including haloes) 
of smaller scales embedded in these sheets, and only 46\% is in haloes.
Thus, no matter how we look at it,
the filaments contain the largest fraction of mass,
and their higher density contrast makes them much easier to detect than
the sheets. 
Several other studies have addressed the large-scale geometry of 
cosmic-web filaments in simulations 
\citep[e.g.,][]{colberg05, hahn07b, hahn09, gay_pichon10, aragon10, noh11},
and how to trace it in the observed distribution of galaxies 
\citep[e.g.,][]{dekel_west85, aragon07b, sousbie11I, sousbie11II}.
\nocite{aragon10s}
Massive galaxies at redshift $z \geq 2$ form in dark-matter haloes
that are much more massive than the typical halo at that time,
hundreds of times larger than the characteristic Press-Schechter mass 
(which is $\sim 3\times10^9\msun$ at $z=2.5$ for the WMAP7 cosmological 
parameters). 
They are therefore ``high-sigma peaks" of the initial density fluctuation 
field, which reside in the nodes of the cosmic web. 
In each halo,
the dark matter flowing along the filaments feeds the dark matter 
halo and fills the entire virial sphere.
The baryons that stream with the dark matter along the filaments into the 
halo virial radius penetrate through the halo and feed the galaxy at the halo 
centre, both as merging galaxies and a smoother gas component 
\citep{bd03,keres05,db06,ocvirk08,dekel09}.
This is illustrated in \fig{proj}, showing the influx density per solid 
angle on the scale of a few virial radii
in two galaxies from the Mare Nostrum cosmological simulation 
\adb{\citep{ocvirk08}}
used in \citet{dekel09} and in the current paper.

\Fig{bird} shows the gas density in the streams that feed the same galaxy 
shown in the right panel of \fig{proj}, galaxy 314 of the Mare Nostrum 
simulation, which has been displayed in more detail in \citet{dekel09}. 
The current map is from a test run using RAMSES with a maximum resolution 
of $60\pc$, higher by more than an order of magnitude than the Mare Nostrum 
resolution. This figure demonstrates that the general stream pattern is not
affected by the resolution, except that the inner structure of the streams 
and the smaller merging galaxies are better resolved.

The efficient feeding of galaxies by streams has a major impact on
the way these galaxies form and evolve \citep{db06, dsc09, cdb10, oser10,
ceverino11, cdg11}.
The streams bring in most of the baryonic mass and angular momentum, 
and are thus responsible for the formation of large rotating discs.
The continuous intense in-streaming allows for high star formation rates (SFR) 
in those discs, in most cases not related to mergers \citep{genzel06, dekel09}.
The streams maintain a high gas fraction that is responsible for a 
violent disc instability with large transient features and giant clumps, 
which drive a rapid mass inflow to the centre
\adb{
\citep[independent of whether the clumps disrupt][]{dsc09,genel12,hopkins12}.
This may eventually lead} to the 
formation of a central bulge with a massive black hole 
\citep{dsc09,cdb10,bournaud11}.
A study of the baryonic stream properties is therefore crucial for a better
understanding of galaxy formation at high redshift.
In addition, since the gas tends to condense in the central, denser 
regions of the more dispersive dark-matter filaments and sheets, 
the gas can also serve as a unique tracer of the cosmic web skeleton in the 
vicinity of its high-sigma peak nodes.

In this paper we use a hydrodynamical cosmological simulation,
the Mare Nostrum simulation, to study the large-scale
properties of the streams feeding massive galaxies 
at high redshift. We also make a preliminary attempt at relating the angular
momentum (AM) of disc galaxies to the AM carried by the streams from the cosmic
web into the halo. 
The outline of the paper is as follows: 
In \se{method} (and \se{MN}) 
we describe the simulation and the method of analysis.
In \se{SP} we study the tendency of the streams to be co-planar
and the first detection of pancakes in cosmological simulations,
to be studied in detail in \citet{hahn11}.
In \se{flux} we investigate the distribution of influx in streams and pancakes,
revealing one dominant stream and a tendency for three major streams.
In \se{AM} we address the angular momentum transport by streams
from outside the halo into the inner disc,
by exploring the degree of alignment between the stream plane and AM at the
virial radius and between them and the disc.
In \se{theory} we discuss the possible origin of the main features of the
cosmic web near a node.
In \se{conc} we summarize our results and discuss them.
We address several technical issues in the appendices. Of particular interest
is \se{axes}, where we evaluate a potential systematic effect in our analysis, 
namely the numerical tendency of alignment 
between the disc and the axes of the simulation grid.

\section{Method}
\label{sec:method}

\subsection{The cosmological simulation}
\label{sec:sim}

The cosmological simulation used in this analysis is the AMR
Horizon Mare Nostrum (MN) galaxy formation simulation \citep{ocvirk08}, 
which utilized the AMR code RAMSES \citep{teyssier02} to simulate the 
dynamics of gas and dark matter in a cosmological box. 
The standard $\Lambda$CDM cosmology is assumed, with $\oml=0.7$, $\omm=0.3$,
$\omb=0.045$, $h=0.7$ and $\sigma_8=0.95$ in a periodic box of side 
$50\hmpc$. 
The dark-matter component is represented by $1024^3$ particles of 
$1.17\times10^7\msun$ each. 
A basic grid of $1024^3$ cells is progressively adapted when the number of 
DM particles in the cell exceeds 8,
\adb{or when the gas mass exceeds 8 times the initial gas mass resolution.}.
The minimum cell size is $1\kpc$ physical.
The large cosmological box and the 1-kpc resolution allows for a reliable
statistical study of the way massive galaxies are fed from the cosmic web 
at the halo scales.  
This resolution allows for identifying massive discs that extend
to $\sim\!10\kpc$, but the disc thickness is barely resolved.
The MN simulation employs at the sub-grid level
physical processes that are relevant for galaxy formation, such as
star formation, supernovae feedback, metal enrichment, metal dependent 
cooling and background UV heating (see appendix \se{MN}). 
However, these processes are followed with a limited accuracy that is dictated
by the limited resolution.

\subsection{Dark Matter haloes and galaxies}
\label{sec:haloes}

We select from the MN snapshot at $z=2.5$ all the 351 dark-matter haloes 
with virial mass in the range $(0.8-3)\times 10^{12}\msun$ 
(serially numbered 50 to 400 in order of decreasing mass). 
This is the mass range of haloes that host the massive star-forming galaxies 
observed at these redshifts \citep{genzel06,genzel08,tacconi08,dekel09},
with a comoving number density of $4\times 10^{-4} \Mpc^{-3}$
for $\Mv > 1.5\times 10^{12}\msun$ at $z=2.5$.
These haloes are expected to be relatively rare ($> 2\sigma$) density peaks 
at the nodes of the cosmic 
web, and are predicted to have shock-heated hot media penetrated by narrow 
cold streams \citep{db06,dekel09}. 

For each halo, we define the centre as the peak of the gas density smoothed
with a Gaussian of standard deviation $5.4 \kpc$.
We then use the gas density and velocity in $256^3$ cells
of a cubic grid about the galaxy centre. 
For a study on the halo scale at the highest resolution, the 
cell side is 1.246 kpc ($=50/2^{\ell}\hmpc$ comoving at adaptive level
$\ell=14$) 
and the corresponding box side is 319 kpc, almost encompassing a $2\Rv$ sphere.
For a study of the greater environment of the halo, we use boxes and cells
twice and four times as large (adaptive levels $\ell=13$ and 12), 
reaching to $\sim 5.5\Rv$.

The gas velocities are corrected to the rest frame of the centre of mass 
of the \adb{cold} gas 
\adb{with $T < 10^5$K}
within a sphere of radius $0.1\Rv \sim 10\kpc$ 
about the density peak,
mimicking the centre of mass of the disc.
\adb{In most cases this correction is smaller than one resolution element
and it has a negligible effect on the results.}
A Hubble flow ($H=254\kms\Mpc^{-1}$ at $z=2.5$) 
is added to the comoving velocities used in the simulation. 
Angular momentum is computed about the position of this centre of mass.
Our results concerning inflows and angular momenta were tested not to
be sensitive to the exact choice of disc centre and rest frame, which can be 
determined at an accuracy level that is comparable to the simulation resolution.

We exclude from the analysis 15 haloes where there is no obvious central galaxy 
in the halo, presumably due to ongoing major mergers. This leaves us with a
sample of 336 haloes.

\subsection{Identification of streams}
\label{sec:streams}

The streams consist in principle of dark matter, stars and gas, 
but our previous attempts
to quantify stream statistics based on the dark matter led to noisy results,
that got worse at smaller radii inside the haloes where the dark matter
is virialized.
Here we identify the streams and study their properties based on the cold gas,
which, due to dissipative processes, traces much more cleanly the filamentary
skeleton of the cosmic web.

The main quantity used at every point $\vec{r}$ on a spherical grid about the
galaxy centre is the 
radial flux density per solid angle,
\be
\frac{d\dot{M}}{d\Omega} = \rho v_{r} r^{2},
\ee
where $\rho$ is the gas density, $v_r$ the radial velocity and $r$ the radial
distance from the galaxy centre. 
We use in the analysis thin spherical shells of one-cell thickness
or the average over several thin shells inside a thick shell.
The scalars given in the cells of the cubic grid are interpolated 
into a spherical grid of $512^2$ equal-area angular cells, with a radial 
thickness of $\sqrt{3}\times50/2^{\ell}\hmpc$, where $\ell$ is the
adaptive level used for the cubic grid that samples that shell 
(see \se{haloes}).

\begin{figure*}
\includegraphics[width=1\textwidth]{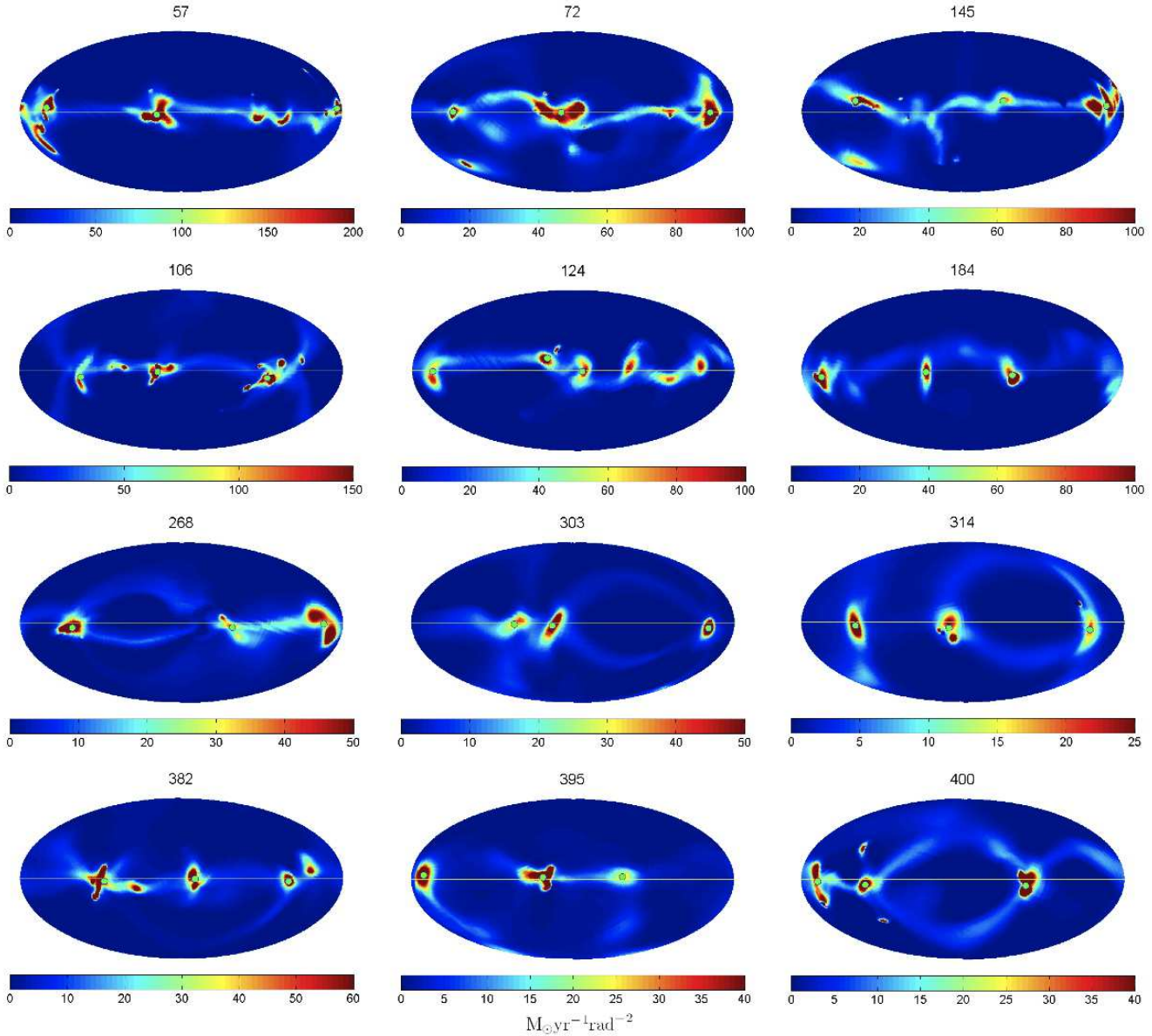}
\caption{
Co-planar inflowing streams and pancakes in a thick shell at $(1-2)\Rv$.
Shown as whole-sky Hammer-Aitoff maps of influx density are twelve different
simulated galaxies in haloes of $\sim\!10^{12}\msun$ at $z=2.5$.
They are selected to represent the many galaxies with at least 3 
streams that almost perfectly lie on a great plane.
The coordinates are rotated such that the best-fit stream plane coincides with
the equator.
The color represents radial influx of gas mass per solid angle.
The centres of the three streams with the highest influx are marked by green
dots.
Typical flux densities in the streams are $(50-150)\syr$
with the higher fluxes valid in the more massive haloes.
Thin pancakes of $\sim 10-15\syr$ are seen between the streams, most frequently
coinciding with the stream plane, but sometimes showing pancake segments that
deviate from the stream plane.
}
\label{fig:planarity_ex}
\end{figure*}

In each shell, we identify streams by first applying a given influx density 
threshold, at a given overdensity above the average influx density in 
the entire shell. 
We then apply an angular friends-of-friends algorithm to the influx 
above the threshold on the spherical shell grid. 
This means that we assign to the same stream all the adjacent grid cells 
with influx density above the threshold
(see \se{flux} for more details). 
For each stream, we record its flux-weighted average angular position on the 
sphere and its total mass inflow rate. 

In our analysis of the streams flowing into the virial radius, 
we sometimes consider 
\adb{the sum of all the}
spherical shells in the radius range $(1-2)\Rv$. 
This stacking improves the statistics 
\adb{by including more information}, and 
\adb{in particular} 
it emphasizes elongated radial streams that stretch over a virial-radius scale. It properly smoothes over
the local fluctuations due to merging galaxies, putting them in the context
of the streams that they belong to.

\begin{figure*}
\includegraphics[width=0.49\textwidth]{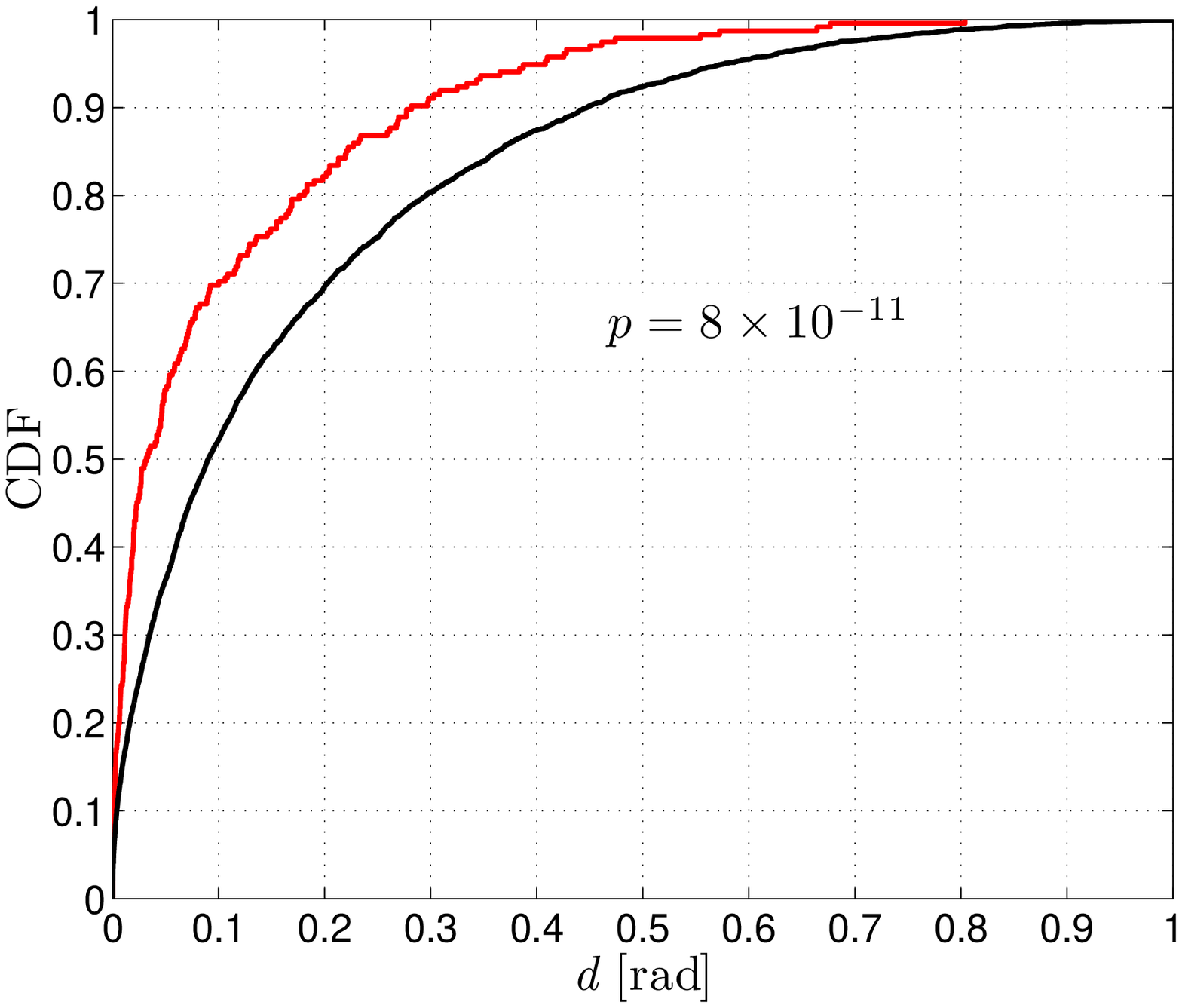}
\includegraphics[width=0.49\textwidth]{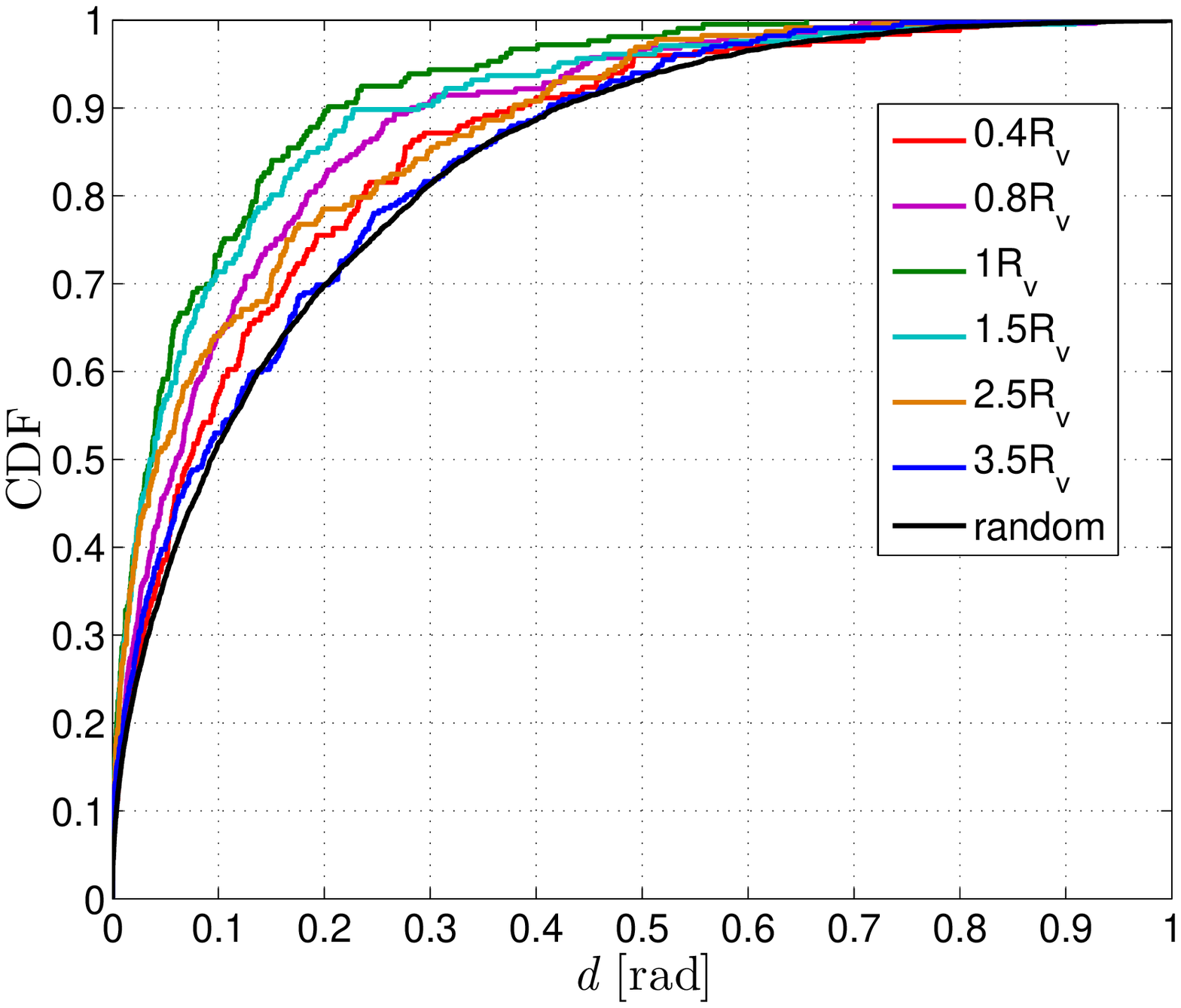}
\caption{
The significance of the stream co-planarity in a shell $(1-2)\Rv$ (left)
and in shells of different radii (right).
Shown is the cumulative probability distribution (CDF) of the rms deviation
of the three largest streams from the best-fit plane, $d$,
based on 235 haloes (color).
It is compared to the same CDF for the null hypothesis of
a random angular distribution of 3 points, based on 5,000 realizations (black).
The same threshold $s_{\rm min}$ is applied to the data and the null model.
A KS test reveals that at $\Rv$ the null hypothesis is strongly rejected
with a p value of $8\times10^{-11}$.
The corresponding p-values for shells from $0.4 \Rv$ to $3.5 \Rv$ are:
$0.12, 7\times10^{-5}, 2\times10^{-12}, 2\times10^{-8}, 4\times10^{-6}, 0.38$.}
\label{fig:main_kstest}
\end{figure*}

\begin{figure*}
\includegraphics[width=0.49\textwidth]{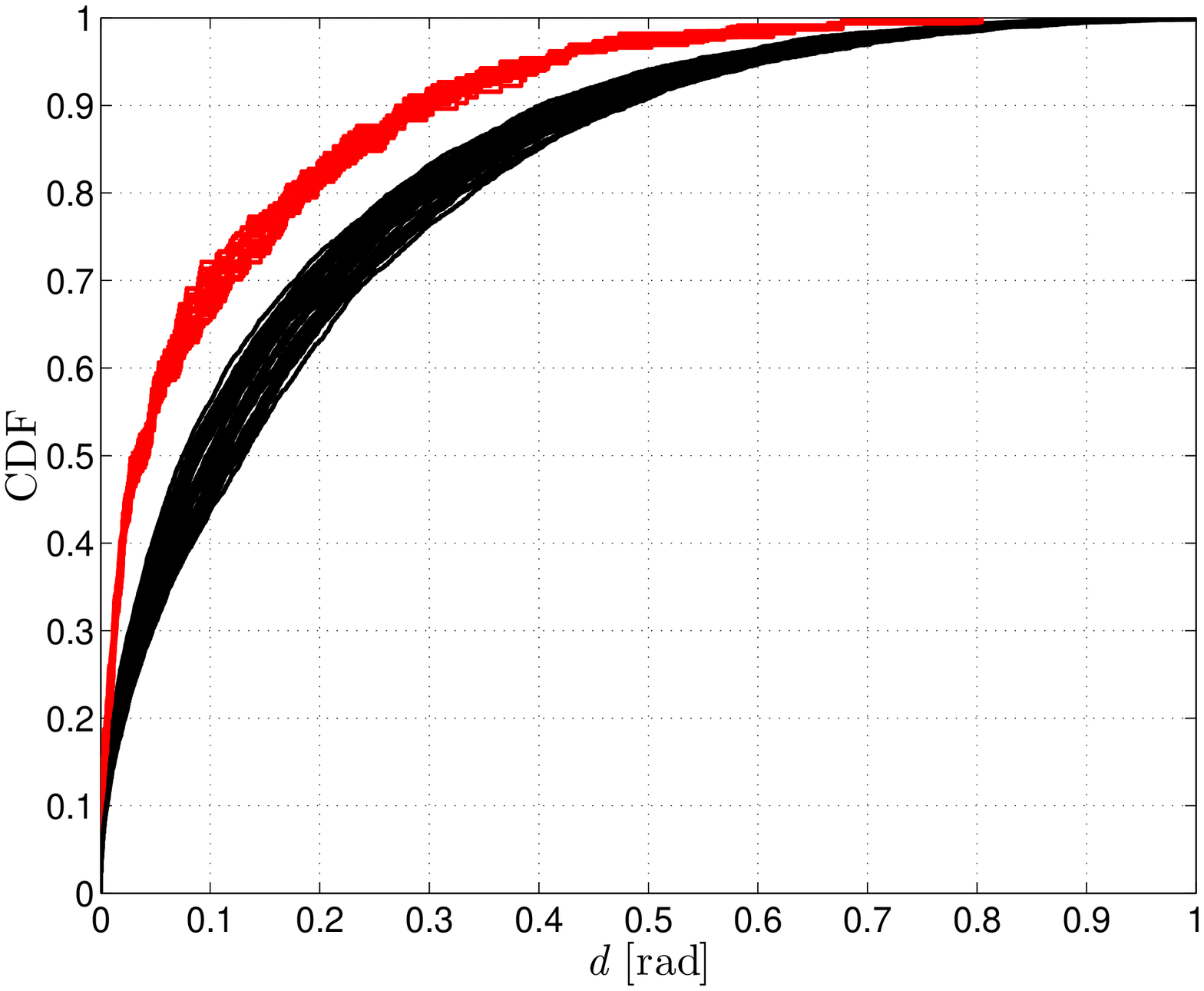}
\includegraphics[width=0.49\textwidth]{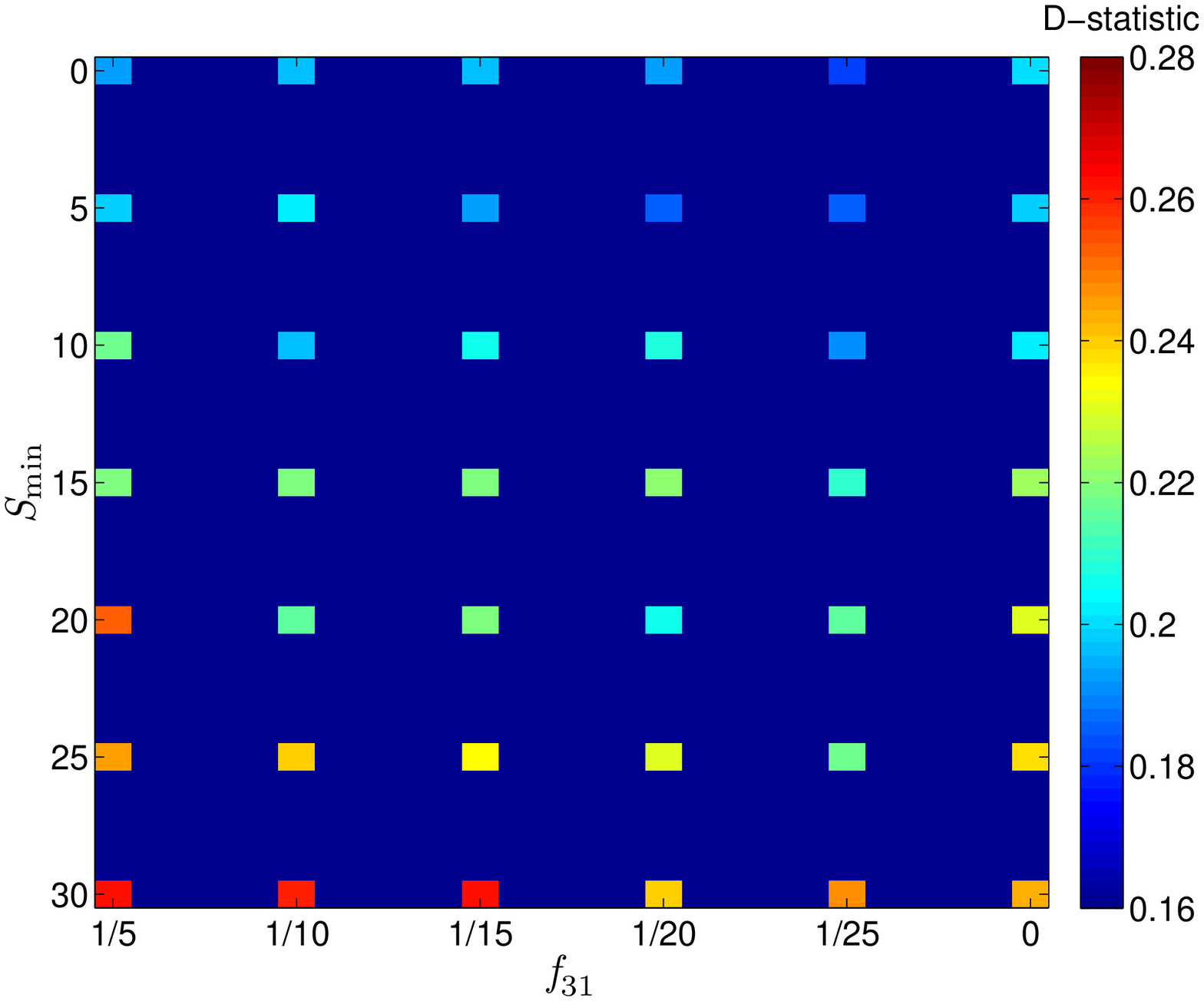}
\caption{
Sensitivity to the procedure of identifying the streams.
{\bf Left:} Shown are the CDFs of $d$ for different values of $s_{\rm min}$
and $f_{31}$, for the simulated haloes (red) and the random-position
null model (blue).
{\bf Right:} The D statistic of the KS-test (in color)
for different combinations of $s_{\rm min}$ and $f_{31}$.
The results are insensitive to the choice of parameters in the given range.
}
\label{fig:robust}
\end{figure*}

\subsection{Numerical Alignment}
\label{sec:numerical}

A potential caveat in our analysis of preferred planes is the numerical 
tendency for artificial alignment of planes of matter with the simulation grid. 
A grid-based Poisson solver, and in particular the hydrodynamical solver,
create non-physical forces along the preferred Cartesian directions of the 
simulation grid, which act to align the mass distribution with the grid 
\citep{hahn10}.
This is especially relevant for the galactic disc plane, which involves 
scales not much larger than the resolution scale, but it would also be 
useful to verify that no artificial alignment propagates to the sheets 
on larger scales 
\adb{(to be defined below)}.
The artificial alignment is expected to be stronger at lower redshifts, where
the discs might have had enough time to relax to the closest grid direction,
so we expect our analysis of massive galaxies at $z=2.5$ to be less 
vulnerable to this numerical effect, despite the 1-kpc resolution.

We address this potential numerical alignment in appendix \se{axes}. 
We find that it is limited to $\sim 20\%$ of the discs.
For cases where the original cosine of the angle between the normal to the
disc and the closest grid axis was $\cos \theta > 0.8$, it has been pushed up
by $\Delta(\cos \theta) \sim 0.08$.
This is a rather small effect, associated with a negligible shift in the mean 
of $\cos\theta$.  This is the level of error that we should assign to any 
measure of alignment between the disc and other planes.
As expected, the stream plane at $\Rv$ does not show any measurable numerical 
alignment with the simulation grid.

\section{Stream co-planarity}
\label{sec:SP}

In the large-scale cosmic web, where the filaments bridge between the nodes 
to form a three-dimensional network, there is no obvious a priori reason for 
the filaments that feed a node to lie in one plane.
We find below that they do tend to lie in a plane that contains the galaxy
centre, which we term ``the stream plane" (SP).
\Fig{planarity_ex} shows the influx in a thick shell at $(1-2)\Rv$
for twelve different simulated galaxies.
The haloes shown were selected from the large sample of haloes 
that have at least three streams (according to our definition of stream number
described below) and they are very well confined to a great plane
(namely a plane that contains the galaxy centre).
The coordinates are rotated such that the best-fit stream plane 
is at the equator.
We find using the goodness of fit measure (described below)
that 60\% of the galaxies fit a plane better than the worst case shown 
in the figure (halo 145).
Examples of galaxies in which the fit to a plane is worse are shown 
in appendix \se{bad}.

\subsection{The best-fit Stream Plane}
\label{sec:fit}

We define the best-fit stream plane and measure the quality of the fit
based on the angular positions of the centres of the three streams with the
largest influx at $(1-2)\Rv$.
One or two streams always lie on a plane that contains the galaxy centre,
so three is the smallest number of streams for which co-planarity is nontrivial.
We thus eliminate from the analysis of co-planarity
the haloes with less than three streams (see below).
We will show that typically more than 90\% of the influx in streams 
is carried by the three leading streams, so using these three for the fit
and ignoring any additional streams is quite sensible.
We limit the fit to a fixed number of streams (3) in order to properly 
compare to a null hypothesis of random angular positions without worrying
about the dependence of the co-planarity on the number of points. 
For a similar reason, when we fit a plane we do not weigh the three streams by 
their individual influxes.

The three streams for the fit of a plane are determined after applying
the two following procedures.
First, if two streams are separated by less than a minimum angular separation 
$s_{\rm min}$, we combine them into one, with the sum of the mass inflow rates
and a new inflow-weighted centre. 
This is in order to avoid confusion with two apparently nearby
streams that may actually be part 
of one stream and may lead to a trivial co-planarity. 
Second, and what turns out to make a stronger effect, 
we eliminate from the analysis haloes in which the ratio of inflow rate 
in the third stream compared to the first stream is smaller than a 
threshold $f_{31}$.
If this ratio is below the threshold, we consider the halo to be of two streams
or less, where the fit to a plane should be trivial, and eliminate it from the
analysis of co-planarity.
These procedures reduce the number of haloes that we use in our statistical
analysis concerning the stream plane from 336 to 235.
The main selection is due to the threshold $f_{31}$, which with $f_{31}=0.1$
filters out 92 of the 101 haloes that are eliminated from the co-planarity
analysis.

\begin{figure}
  \centering
\includegraphics[width=0.49\textwidth]{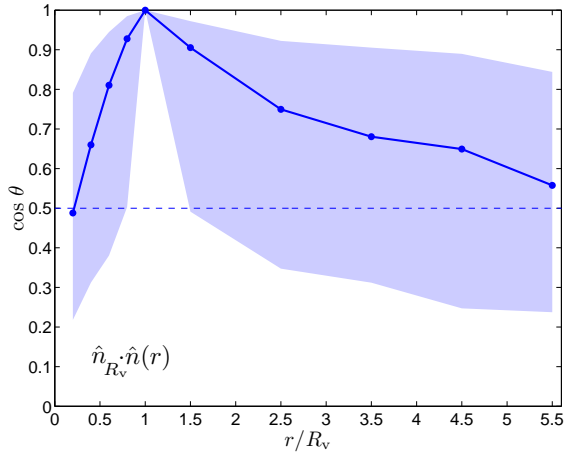}
\caption{
Extension of the streams plane. Shown is the median of the cosine of the angle
between the SP at $1\Rv$ and the SP at different radii from
$0.2\Rv$ to $5.5\Rv$ compared to a random distribution in which the median 
is 0.5. It shows 
a significant alignment from 
\adb{$0.5\Rv$}
to $2.5\Rv$ and a weak alignment
for shells inside $0.5\Rv$ and in the range $2.5-5.5\Rv$.
}
\label{fig:plane_cor}
\end{figure}

The deviation of the three streams at a given concentric shell
from a given great plane is defined by the sum in quadrature of the angular 
distances between the stream centre angular positions and the great circle of 
intersection between the plane and the shell.
The best-fit plane is determined by minimizing this sum,
and the minimum value, $d$ in radians, 
is used as a measure of the deviation from a plane.

In appendix \se{strip}, we describe a different algorithm for
defining the stream plane, which maximizes the inflow rate through a belt 
of width $\pm \pi/9$ radians about a great circle on the spherical shell.
We find that in the vast majority of the haloes, at $\Rv$,
the stream planes defined by 
the two methods practically coincide, with a median of $\cos\theta = 0.96$ 
for the angle between the planes as defined by the two methods (\fig{strip}).

\subsection{The significance of the co-planarity}
\label{sec:coplanarity}

The statistical significance of the stream co-planarity is
evaluated by comparing the data from the simulations
to a null hypothesis where the
streams represent three random points on the sky,
obeying the same selection criteria based on $s_{\rm min}$ and $f_{31}$.
We use a Kolmogorov-Smirnov (KS) test to compare the cumulative
distribution functions (CDF) of the deviation from a plane, $d$,
in the sample of simulated haloes compared to the null hypothesis.
The KS-test allows to determine the statistical likelihood (p-value)
that the two data sets are drawn form the same distribution.
\adb{For the purpose of identifying the centers of the streams,}
the fiducial values for the selection parameters are chosen to be
an influx density threshold for the stream identification of 
5 times the mean influx density
in the shell, and then $s_{\rm min}=15^{\circ}$ and $f_{31}=0.1$.
\adb{The choice of an overdensity threshold of 5 helps sharpening the 
definition of the stream centers, but the results for thresholds of 3 and 
even 2 do not make a significant difference to the resultant co-planarity.}
\Fig{main_kstest} (left) shows the two CDFs, revealing a significant difference
between the simulated data and the random null model.
With a sample of 235 simulated haloes, the maximum vertical separation
between the CDFs is $D=0.234$, implying that the null hypothesis is
strongly rejected with a p value of $8\times10^{-11}$.

\Fig{robust} (left) shows the same CDFs for different values of
$s_{\rm min}$ in the range $0-30^{\circ}$ and $f_{31}$ in the range $0-0.2$.
The value of $D$ for each combination of values for these parameters
is also shown in \fig{robust} (right).
The small variations in the KS-test results demonstrate that the significance
of the stream co-planarity is not sensitive to the exact choice of values
for the parameters used to select the streams for the fit.

\subsection{Spatial extent of the streams plane}
\label{sec:extension}

In order to address the spatial extent of the stream plane
both outside the virial radius and inside the halo, we first
inspect the alignment between the SP at $\Rv$
and the SP at other radii $r$ in the range $0.2\Rv$ to $5.5\Rv$.
\adb{The SP is identified using the same selection criteria at all radii.}
\Fig{plane_cor} shows the median and the 68\% percentiles
for the cosine of the angle between these planes
as a function of $r$, compared to a random distribution where the median is
$0.5$.
\adb{The average cosine is above $0.75$ in the range $0.5-2.5\Rv$,
indicating a significant alignment of the SP in this range.
Note that $\Rv$ itself is not a special physical radius, as the virial 
shock and the dark-matter virialization radius could actually be anywhere in 
this range.
The figure would therefore be qualitatively similar had we anchored 
the curve at some other radius slightly different from $\Rv$.}

Outside the virial radius, the SP extends to beyond $5\Rv$.
KS-tests yield p values that range from $3\times10^{-49}$ at $r=1.5\Rv$,
through $5\times10^{-11}$ at $r=3.5\Rv$, to $6\times10^{-4}$ at $r=5.5\Rv$. 
This plane connects the neighborhood of the halo to the global cosmic web.

Inside the halo, the SP penetrates to $\sim 0.4 \Rv$, 
with the KS p values ranging from $10^{-53}$ at 
$r=0.8\Rv$ to $4\times10^{-7}$ at $r=0.4\Rv$. 
However, at $r=0.2\Rv$, the SP no longer correlates with the SP at $\Rv$,
with a KS $p=0.92$.  This argues that the stream plane practically disappears 
at $r=0.2\Rv$ and inside it, the $20 \kpc$ vicinity of the disc.
We will get back to this in the discussion of angular momentum in \se{AM}.

A complementary way to examine the extension of the SP is by evaluating 
the goodness of fit to a plane in each radius. 
We apply a KS test similar to \se{coplanarity}
to the streams at radii from $0.4\Rv$ to $3.5\Rv$. 
\Fig{main_kstest} (right)
shows the CDFs for the different radii, and the caption
quotes the corresponding KS p values.
We see that the co-planarity is highly significant for radii from $0.8\Rv$
to $2.5\Rv$, and it becomes much less significant both at $r\leq 0.4\Rv$  
and at $r \geq 3.5\Rv$.

We note that the apparent disappearance of the stream plane 
in the inner halo may partly be attributed to the method of 
identification of streams, which becomes less reliable at small radii,
where the streams occupy a larger angular area and the velocity field 
becomes much more complex (see \se{messy} and \fig{messy} below).

\begin{figure*}
\includegraphics[width=1\textwidth]{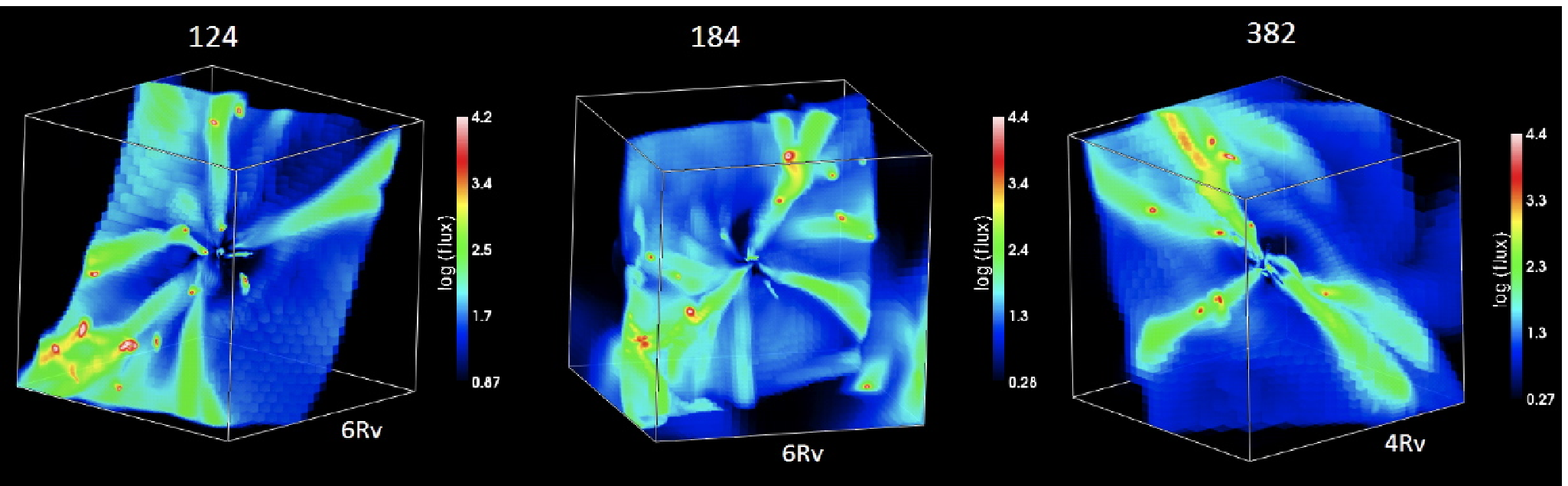}
\includegraphics[width=1\textwidth]{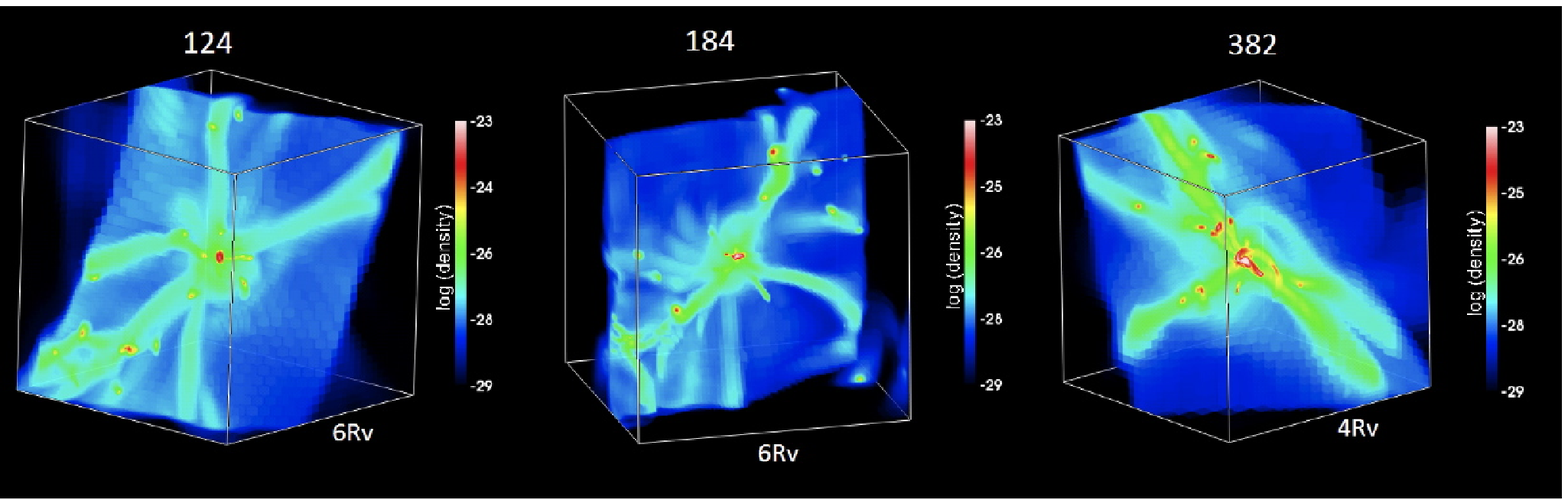}
\caption{
Cold streams embedded in pancakes, flowing into three different
simulated galaxies.  The box side is marked in each case.
\adb{No temperature threshold is applied.}
{\bf Top:} log influx density \adb{(in $\syr$)}.
{\bf Bottom:} log gas density \adb{(in ${\rm g}\,{\rm cm}^{-3})$}.
The planar pancakes and elongated streams become apparent in this ray-casting
maximum-intensity projection, which shows the maximum
intensity volumetric pixel along the line of sight.
}
\label{fig:pancakes3d}
\end{figure*}

\begin{figure*}
\includegraphics[width=1\textwidth]{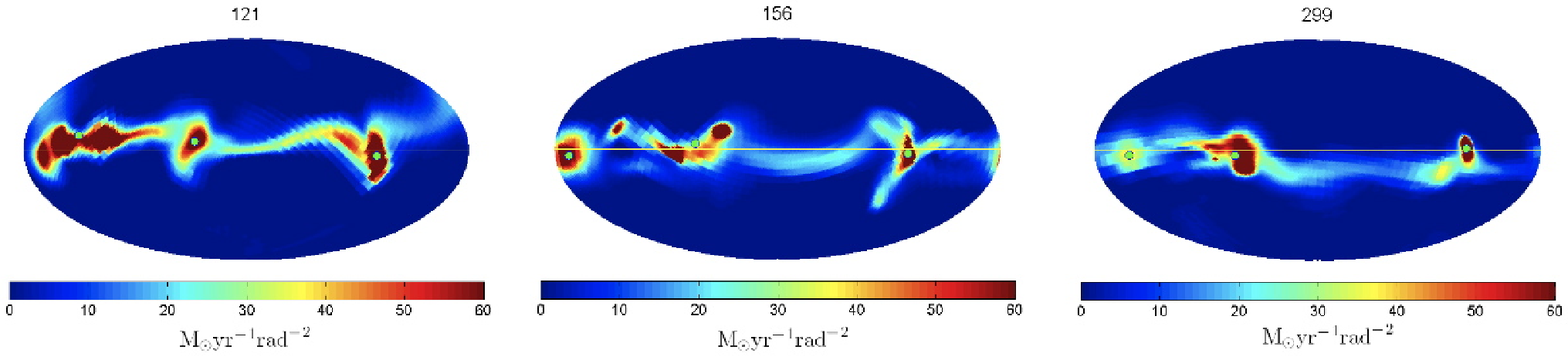}
\includegraphics[width=1\textwidth]{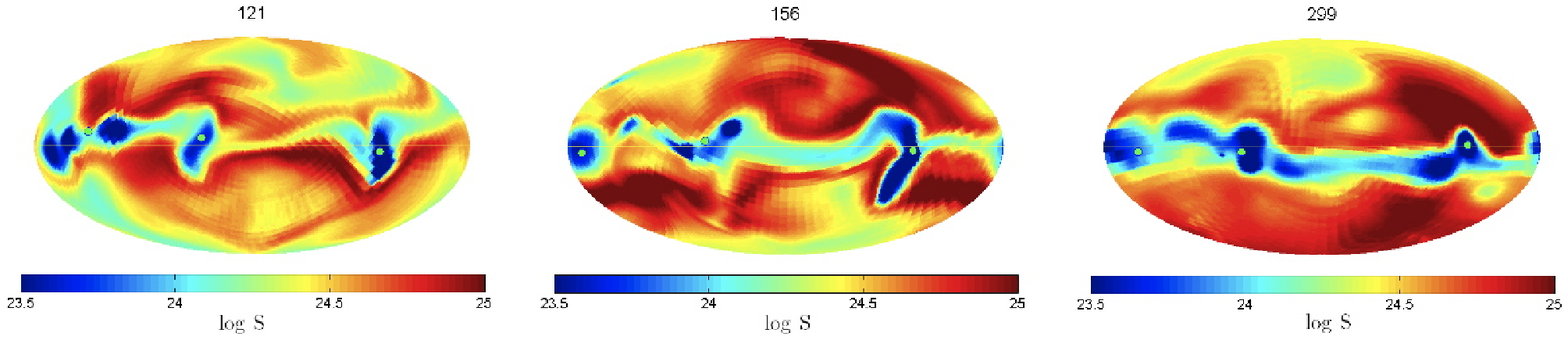}
\caption{
Pancakes at $\Rv$.
Shown are maps of influx density (top) and log entropy $S=T\rho^{-2/3}$ 
(bottom) in three simulated galaxies with dominant pancakes.
\adb{No temperature threshold is applied.}
The coordinates are rotated such that the maximum influx is about the 
equatorial plane (appendix \se{strip}).
More than 90\% of the influx is in the $\pm \pi/9$ equatorial belt.
The pancake and streams are of low entropy, bounded by high-entropy
regions indicative of the planar shocks that resulted from the inflow
of gas from the voids into the pancakes.
}
\label{fig:pancakes_flux_entropy}
\end{figure*}

\subsection{Pancakes}
\label{sec:pancakes}

While the qualitative tendency of the streams to be co-planar has been noticed  
already in \citet{dekel09}, the Hammer-Aitoff projections at and outside $\Rv$, 
like the ones shown for a sample of galaxies in \fig{planarity_ex}, also reveal 
thin sheets of inflowing gas, typically filling the same plane defined 
by the streams, and sometimes hinting to additional planes or plane segments. 
This is a low-entropy gas, inflowing toward the central galaxy
with a typical influx density $\sim\!10 \syr$, as opposed to 
$\sim 100 \syr$ in the streams.
\Fig{pancakes3d} highlights these sheets in three-dimensional images
of gas influx density and gas density in three different haloes.
In each case, the intense streams are embedded in a well-defined sheet of gas 
with influx density just above the threshold, extending to $5\Rv$ and beyond.
These are the pancakes envisioned by Zel'dovich \citep{zeldovich70}, 
showing up for the first time so clearly in the gas distribution of
cosmological simulations, as they are low-density features that are 
normally overwhelmed by the denser, more massive filaments.

\Fig{pancakes_flux_entropy} shows three examples of haloes in which the
high influx density is largely confined to one obvious pancake and its embedded
streams.
In these cases the belt of $\pm \pi/9$ about the best-fit plane (\se{strip}),
encompassing a third of the shell area, contains more than 90\% of the total
influx at $\Rv$.
The entropy maps, where log entropy is measured from the gas temperature
and density as $S=T\rho^{-2/3}$,
demonstrate that the pancakes are regions of low entropy,
hosting the denser, cold streams.
The pancakes are bounded from above and below by regions of higher
entropy, indicative of planar shocks that resulted form the gas flowing
from the voids into the pancakes.
The gas pancakes between the streams are explored in more detail in a 
companion paper using cosmological simulations of higher resolution
\citep{hahn11}.

\begin{figure}
\includegraphics[width=0.5\textwidth]{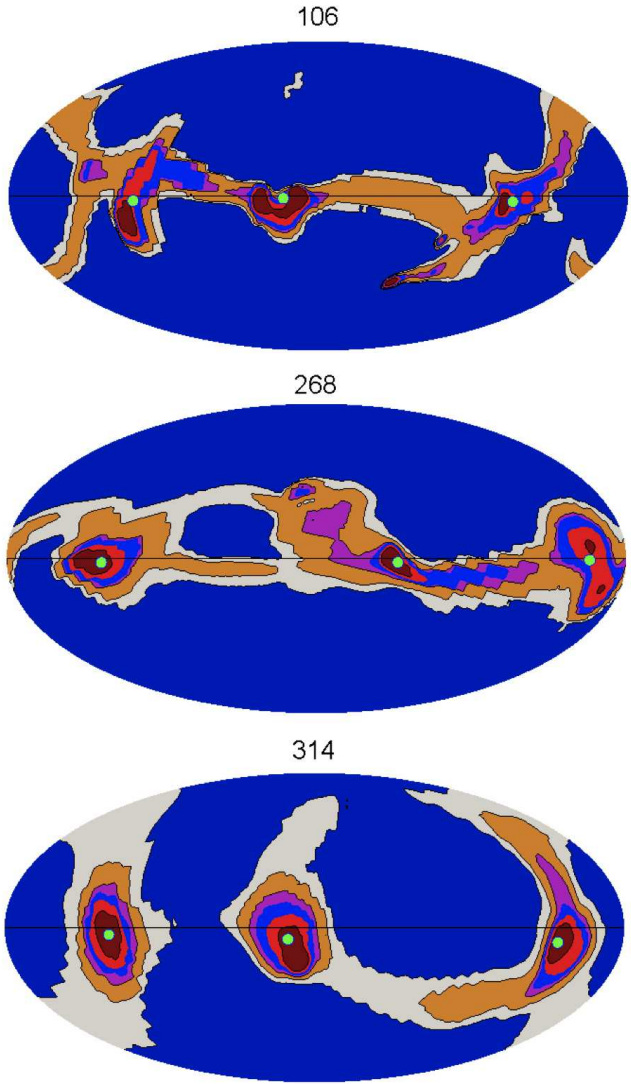}
\caption{
Influx density in streams and pancakes.
Shown is influx density in Hammer-Aitoff maps of three haloes at $\Rv$.
The contours refer to iso-influx-overdensity with respect to the mean
influx density in the shell, at levels of 0.5 (grey), 1 (orange), 2 (purple),
3 (blue), 5 (red), and 8 (dark red).
The pancakes are confined to influx overdensities in the range from below 1 to
about 2.
The streams are defined by influx overdensities of 2 (purple) and above.
The mean values for the three examples are: $9.8, 5.8, 2.9 \syr$.
See \fig{distrib1} and \fig{area}.
}
\label{fig:aitoff_threshold}
\end{figure}

\section{Influx in 3 streams and pancakes}
\label{sec:flux}

We now address the distribution of influx among the streams
and in the pancakes. We explore the effective number of dominant
streams, and justify our focus on the three dominant streams in the
analysis of the stream plane in the previous section.

\Fig{aitoff_threshold} shows Hammer-Aitoff maps of influx overdensity
with respect to the mean for each of three galaxies (\se{streams}),
with an emphasis on the influx density levels associated with the streams
and the pancakes.
One can see that the streams are typically separated out for thresholds 
of $\sim 2$ times the mean influx density. Higher levels of influx overdensity
define the central regions of the streams. We therefore use an overdensity
threshold of 3 or 5
to define the stream centres. The pancakes are typically confined to 
influx overdensities in the range 0.5 to 2.
The small angular area contained within an influx overdensity of order unity
demonstrates the highly anisotropic pattern of cold gas streaming 
\citep[see also][]{aubert04}.

\begin{figure*}
\includegraphics[width=0.49\textwidth]
{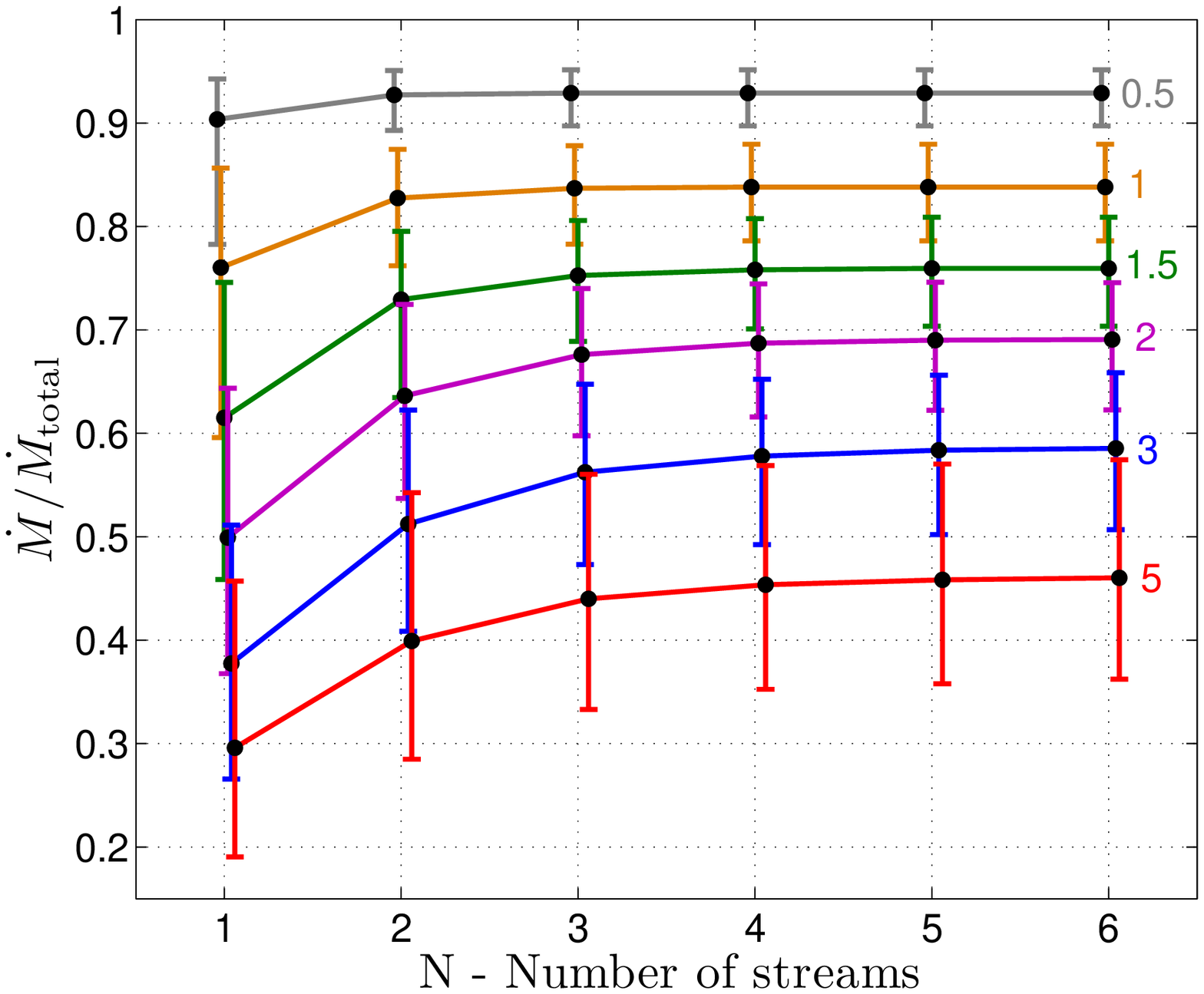}
\includegraphics[width=0.49\textwidth]{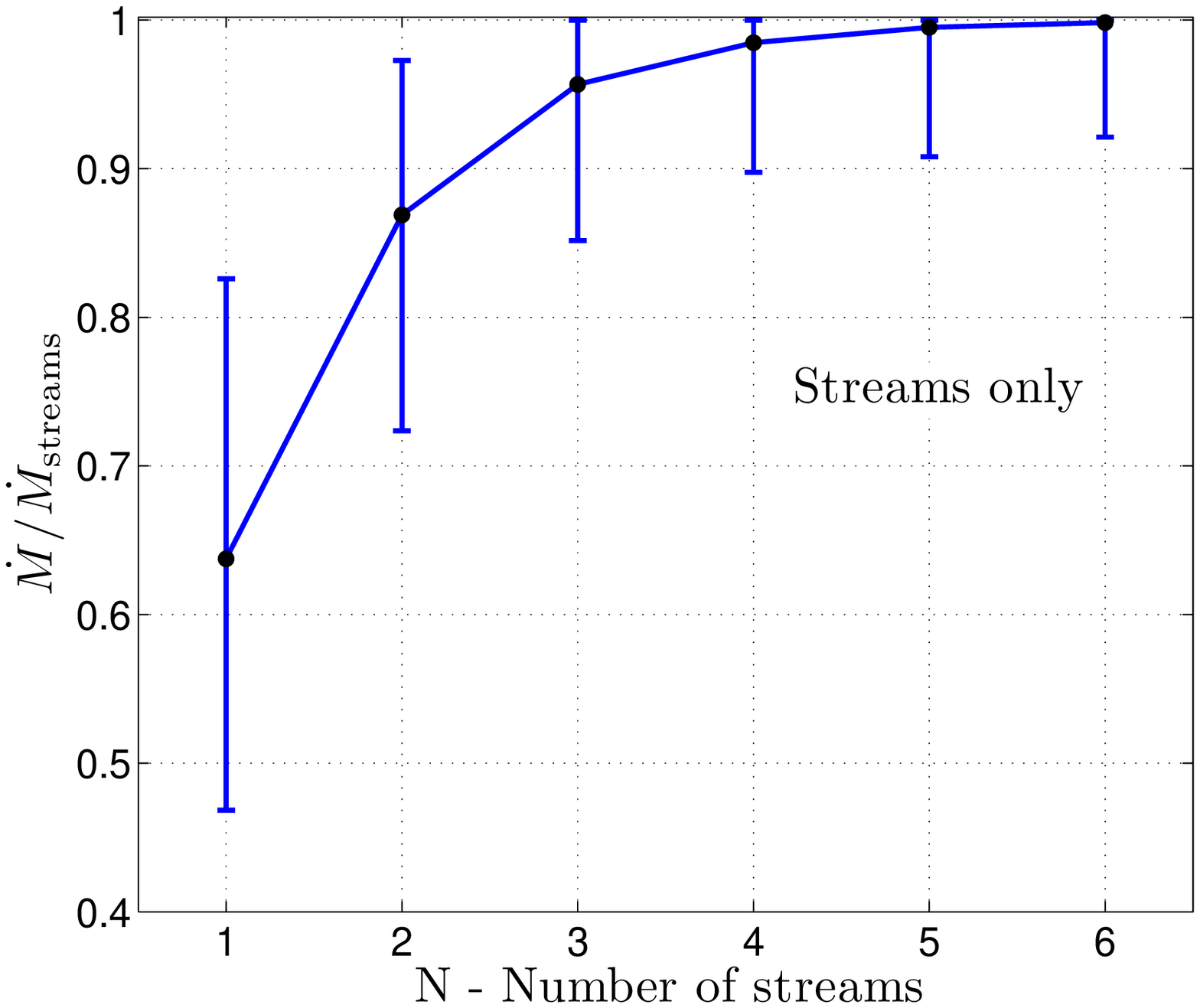}
\caption{
Fractional contribution of the first $N$ streams to the total influx at
$\Rv$ (left) and to the total influx in streams (right).
Shown are the averages and standard deviations over the sample of
336 simulated haloes.
The left panel compares different thresholds of influx overdensity.
For example, with a threshold at an influx overdensity of 2,
the first 3 streams carry on average 68\% of the total influx.
In the right panel the streams are defined by an influx overdensity of 3.
The first stream carries on average 64\% of the influx in streams,
and the first three streams carry 95\%.
}
\label{fig:distrib1}
\end{figure*}

Once the streams are identified for a given influx overdensity threshold,
they are rank ordered by influx, from high to low.
\Fig{distrib1} (left) shows the influx in the first $N$ streams
relative to the total influx in the shell at $\Rv$, for different values
of the influx overdensity threshold, averaged over all 336 haloes.
As described in \se{coplanarity}, nearby streams have been merged based on
$\emph{s}_{min}=15^\circ$,
and streams with influx much below the influx of the first stream were
eliminated based on $\emph{f}_{31}=0.1$,
but here the analysis was not restricted to haloes of three streams.
With the lowest influx overdensity threshold, 0.5 of the mean, the influx
is effectively in a single meta-stream, or a pancake, that carries more
than 90\% of the total influx in that shell.
With an influx overdensity of 2 times the mean,
for which the streams are already well
separated, the first stream carries on average
49\% of the total flux, the first three streams carry 68\%, and all the
streams carry 70\%.
\adb{The threshold overdensity of 2 is useful for capturing most of the
influx in the streams while avoiding most of the off stream influx in the 
pancakes.}
In $\sim 90\%$ of the haloes the dominant stream
is the densest stream, and in $67\%$ of the haloes the dominant stream
has the highest inflow velocity.
In most cases the densest stream is also the one with the highest velocity,
but in $30\%$ of the haloes the densest stream does not have the highest
inflow velocity.
Once the streams are separated, namely for thresholds of 2 and above,
the shape of the curves seem to be rather independent of threshold, 
indicating that the distribution
of flux in the streams relative to each other is robust, while the total 
influx in streams is obviously a decreasing function of the threshold level.

\begin{figure*}
\includegraphics[width=0.49\textwidth]{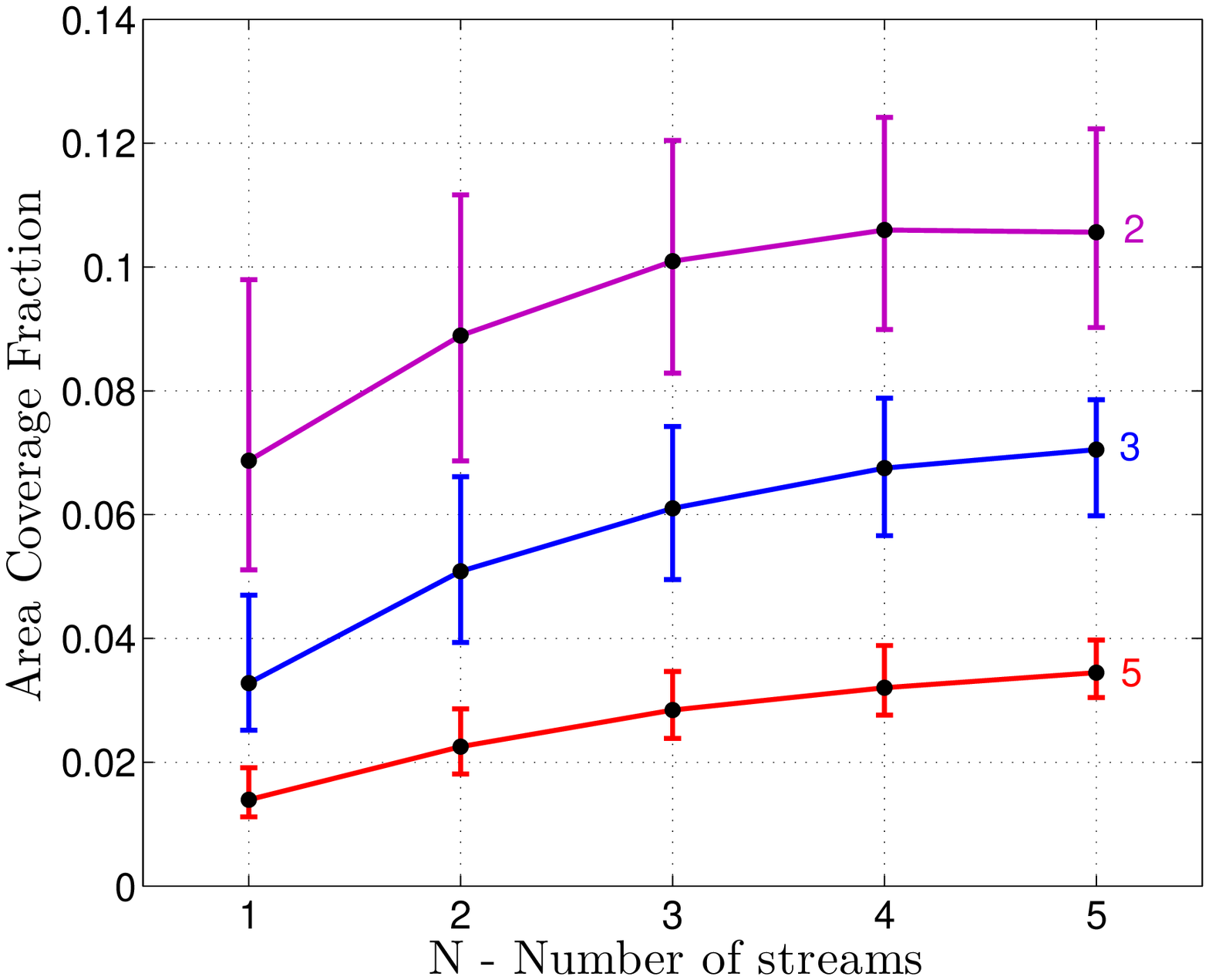}
\includegraphics[width=0.49\textwidth]{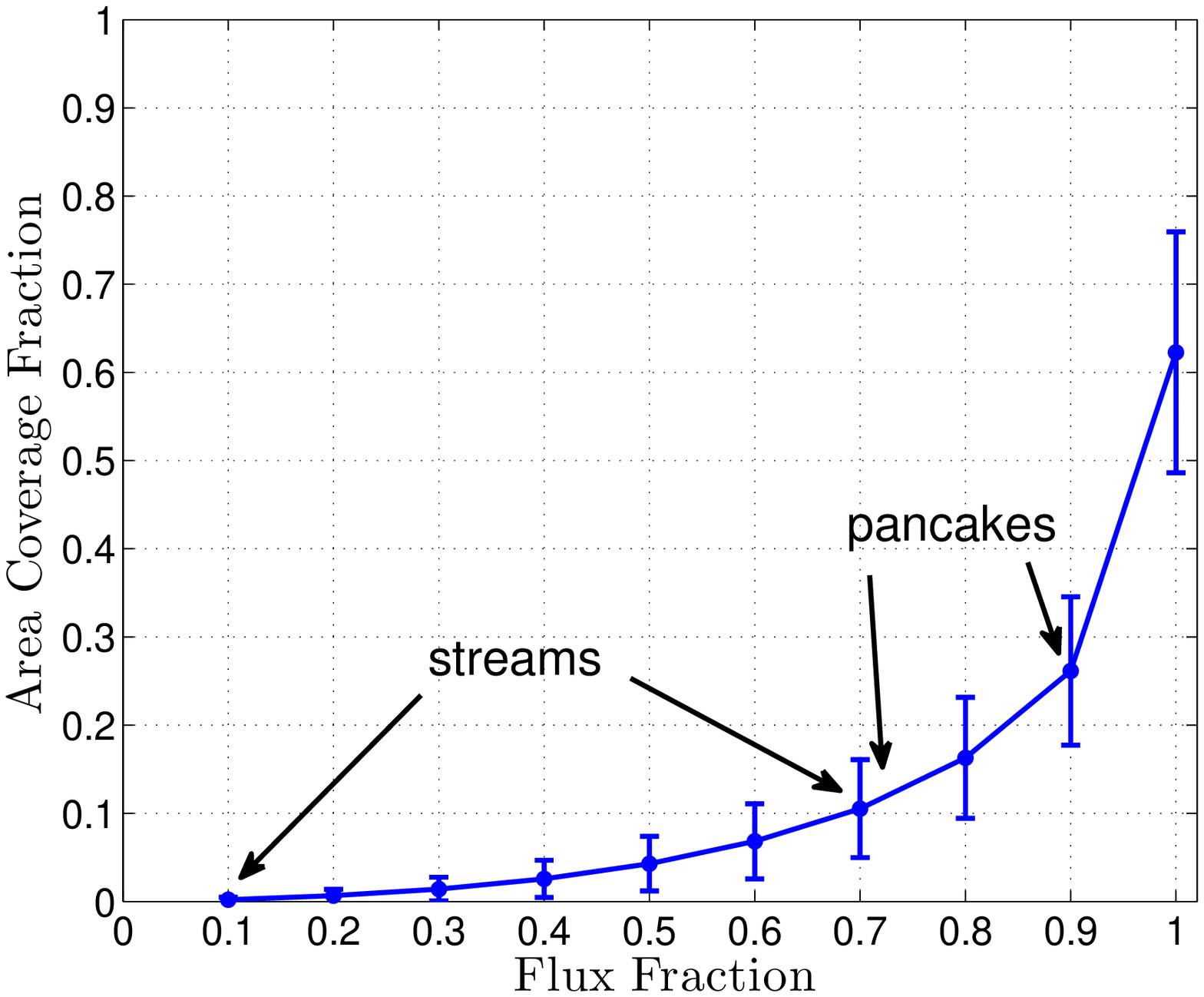}
\caption{
Fractional area coverage by streams and pancakes at $\Rv$ averaged over
the 336 haloes.
{\bf Left:} area covered by the first $N$ streams for 
flux overdensity thresholds 2,3,5.
With a threshold overdensity of 2,
the first 3 streams have an averaging covering fraction of 10\%.
{\bf Right:} area coverage as a function of the corresponding
fractional influx.
The threshold of influx overdensity is varying along the curve,
from $\gg 1$ to $2$ for streams, and from $2$ to $0.5$ for pancakes.
The streams carry on average about 70\% of the influx and cover 10\%
of the angular area.
The pancakes bring in about 20\% of the influx in 20\% of the area.
}
\label{fig:area}
\end{figure*}

\begin{figure*}
\centering
\includegraphics[width=1\textwidth]{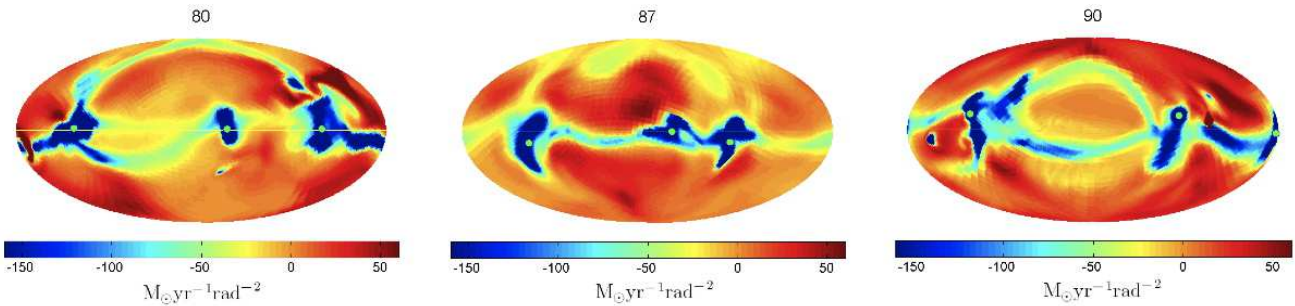}
\caption{Outflowing versus inflowing gas at $\Rv$.
Shown are Hammer-Aitoff projections of the flux density per solid angle
for three haloes, outflowing (red) and inflowing
(yellow to blue) gas. While the inflow is confined to narrow streams and
thin pancakes, the outflows cover $\sim 50\%$ of the shell.}
\label{fig:outflow}
\end{figure*}

\begin{figure*}
 \centering
\includegraphics[width=1.0\textwidth]{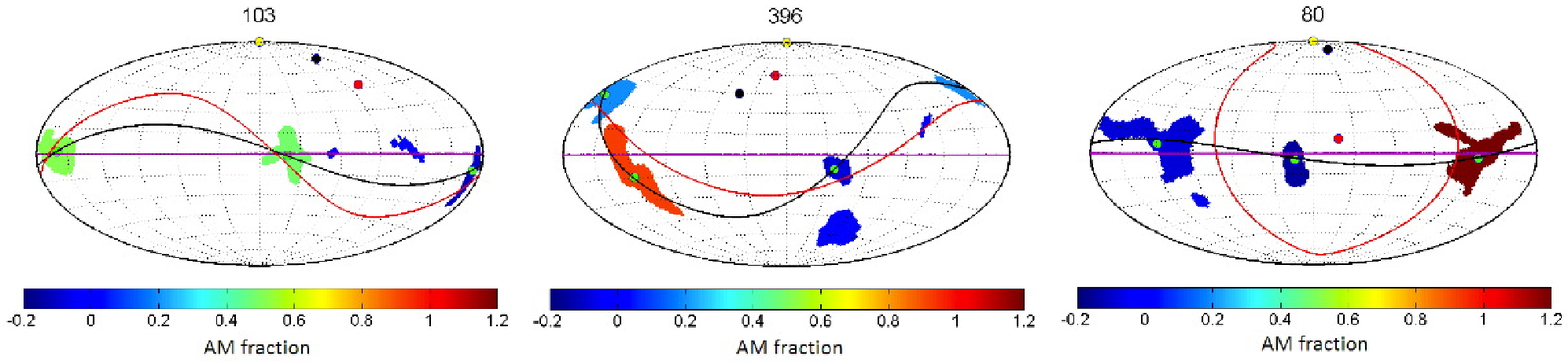}
\includegraphics[width=1.0\textwidth]{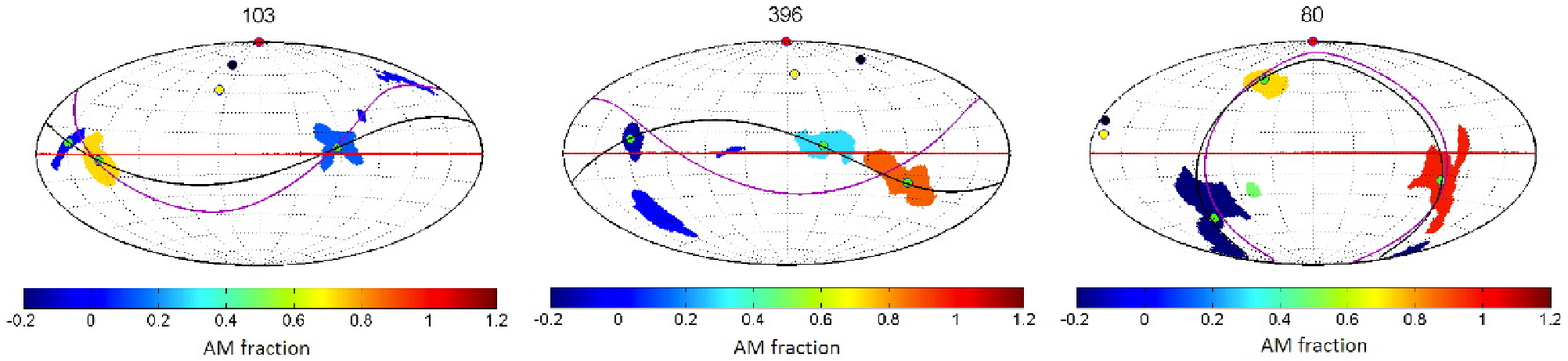}
\caption{
AM fraction in the streams at $\Rv$ shown in a Hammer-Aitoff projection.
The three columns display three galaxies.
Only cells with $V_r<-0.5\Vv$ are included.
Marked by a line and a small circle of the same color are three planes
and the normals to them:
The AM at $\Rv$ (purple), the stream plane (black), and the disc AM (red).
{\bf Top:} Fraction relative to the total AM at $\Rv$, and rotated such that
the AM plane at $\Rv$ is at the equator.
{\bf Bottom:} Referring to the AM component along the
direction of the disc AM, and rotated such that the disc plane is at the
equator. Between the two opposite orientations of the normal to the stream 
plane we show the one closest to the AM vectors.
}
\label{fig:aitoff_r1}
\end{figure*}

\Fig{area} (left) shows the fraction of the shell area covered by the first 
$N$ streams. 
For an influx overdensity of 2, the covering fraction by streams is about 10\%. 
The area covered by the denser parts of the streams
that carry more than half the total influx, defined by an influx overdensity of
4, is on average 5\%.
This is comparable to the area coverage estimated for Lyman-alpha absorption
by cold streams from a central source, using simulations of higher resolution
\citep{fumagalli11,goerdt11}. 

\Fig{area} (right)
shows the fractional angular area covered by streams and pancakes
as a function of the fraction of influx carried by these streams and pancakes,
at $\Rv$.
The curve is constructed by varying the influx overdensity threshold (which is 
not explicit in the plot).
The mean and standard deviation of the area coverage factor at a given flux
fraction, over the 336 haloes, are shown.
We see that on average 70\% of the influx, at the highest influx overdensities,
is limited to about 10\% of the angular area --- this is the influx in
streams, as defined with a flux density threshold of 1.5-2.
At lower influx overdensities, the following 20\% of the influx is covering 
about 18\% of the area, and can be associated with the pancakes.
The final 10\% of the influx is spread over about 30\% of the area,
mostly representing low-velocity infall not directly associated with the
streams, that is unlikely to ever reach the inner disc.

The gas in 38\% of the area is outflowing. 
\Fig{outflow} shows via typical examples how the outflows find their way
trough the broad dilute areas between the inflowing narrow streams and 
thin pancakes.

\Fig{distrib1} (right)
shows the influx in the first $N$ streams relative to
the total influx in streams, averaged over the $\Rv$ shell for an influx 
overdensity threshold of 3 times the mean.
We first note that the first stream is dominant --- it typically carries
50-80\% of the flux in streams.
Then we see that on average 95\% of the stream flux is carried by the first
3 streams (typically ranging from 85\% to 100\%). 
We learn that in the typical halo, for practical purposes, the influx is 
carried by {\it three dominant streams\,} 
with more than half the influx in {\it one dominant stream}.
These estimates are robust --- as seen in \fig{distrib1}, they are not too
sensitive to the influx overdensity threshold chosen in the identification 
of the streams.
 
Earlier related estimates of the effective number of streams can be found
in \citet{colberg05}, \citet{pichon10}, \citet{aragon10} and \citet{noh11}.  
\adb{
\citet{colberg05} studied the dark matter filaments connecting pairs of haloes
more massive than $10^{14}\msun$ on scales of $\sim 20\hmpc$.
They find between one to four filaments per halo, consistent with our findings.
They also find a marginal tendency for an increase in the number of filaments 
for more massive haloes.
\citet{aragon10} obtained similar results for the connectivity of clusters more massive than $10^{14}\msun$
with an average number of filaments in the range 2-5 depending on the mass of the cluster.
}

Being fed by co-planar streams, and having three major streams, 
of which one is dominant,
are not obvious properties of nodes in a three-dimensional network. 
We discuss in section \se{theory} tentative ideas concerning the origin of 
these features,
but it largely remains an open theoretical question.

\section{Stream Plane vs.~Disc: Angular Momentum}
\label{sec:AM}

The cosmic-web streams provide the gas for the buildup of a disc galaxy 
at the centre of the dark halo. The study of the stream properties
provides vital information concerning the process of disc buildup, 
and especially the growth of the angular momentum (AM) that governs
the disc size and structure.
The unique geometrical structure of the in-streaming may help us
visualize and better understand the AM evolution, which is otherwise
quantified in a more abstract way by the tidal-torque theory 
\citep{white84}.
The basic idea is that a small transverse velocity of the inflowing stream 
at a large distance results in a non-negligible impact parameter relative
to the disc centre, which is associated with a large AM that is being 
transported with the stream into the galaxy \citep[e.g.,][]{pichon11}.

\begin{figure*}
\centering
 \includegraphics[width=0.33\textwidth]{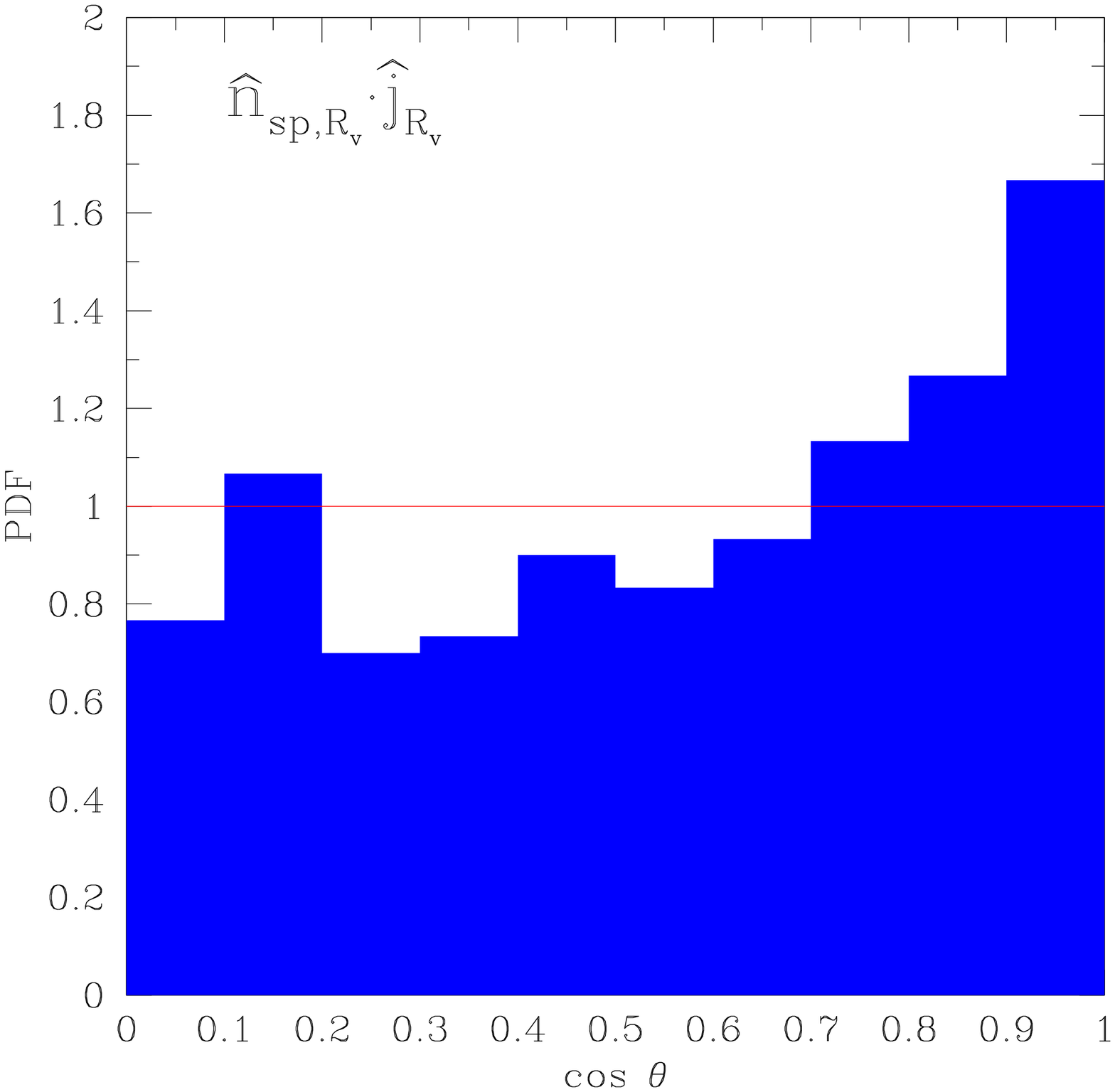}
 \includegraphics[width=0.33\textwidth]{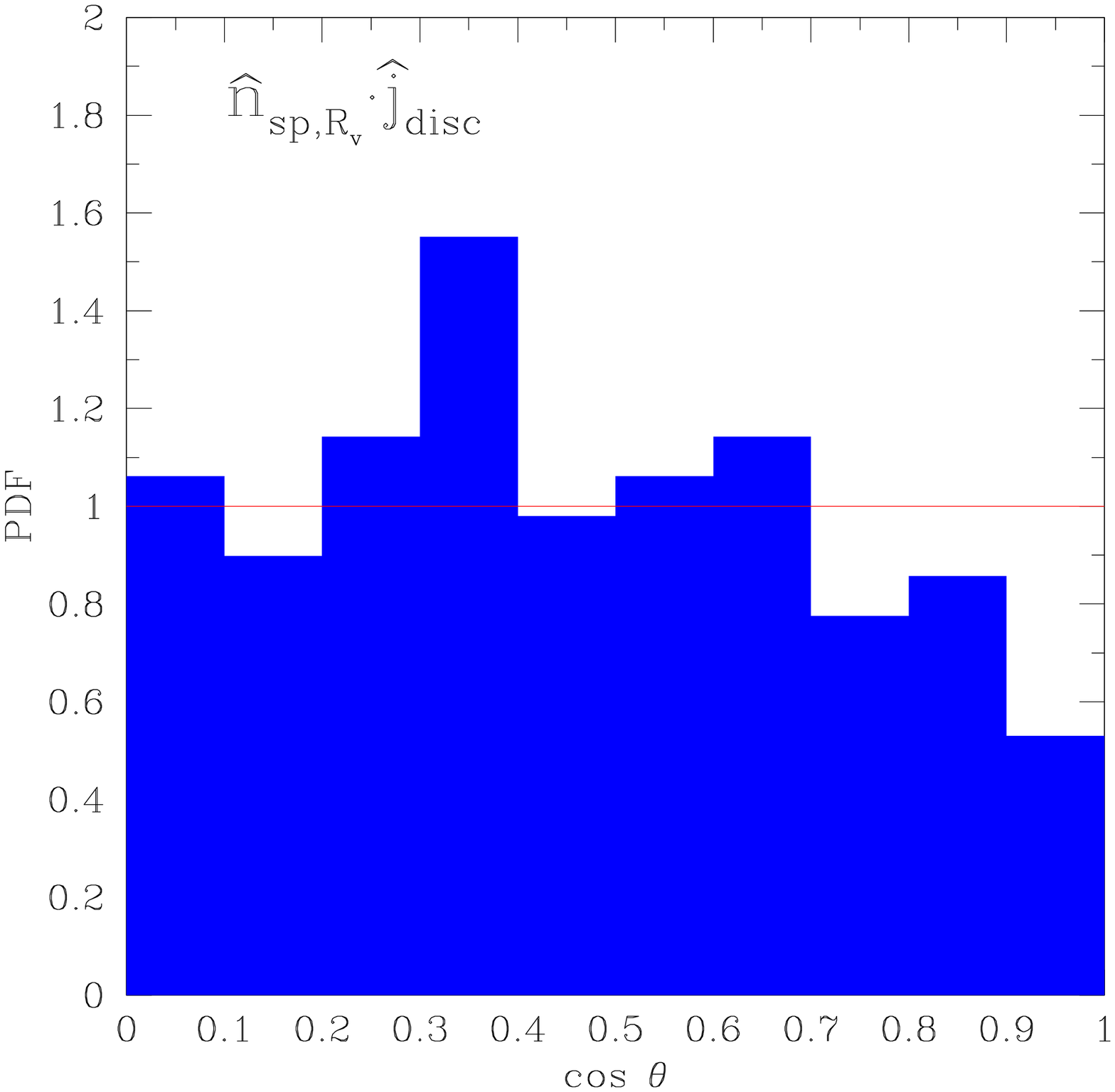}
 \includegraphics[width=0.33\textwidth]{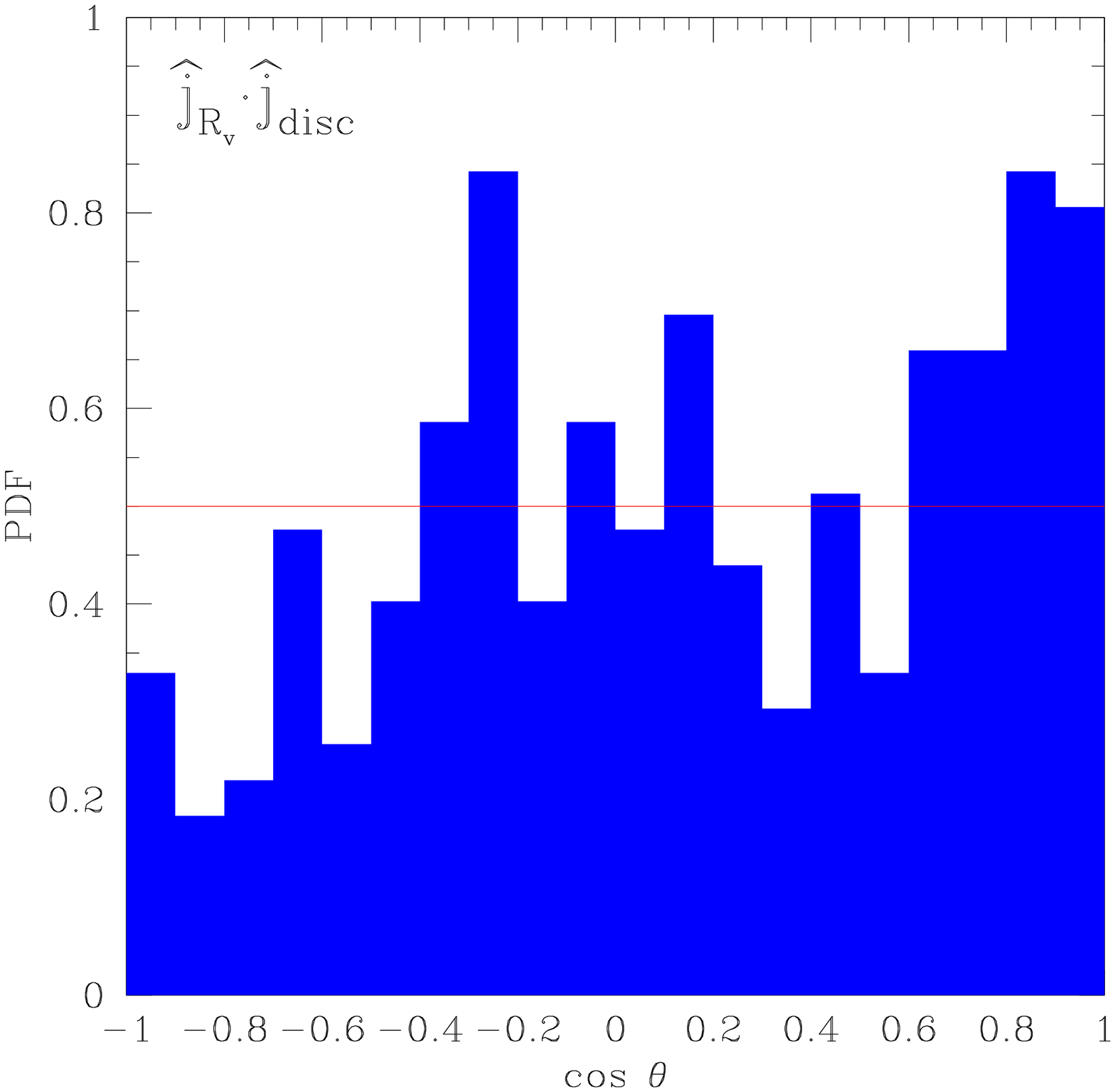}
\caption{PDF of the cosine of the angle between pairs of planes.
{\bf Left:} The stream plane and the AM at $\Rv$, with
$\la cos(\theta)\ra =0.56$, median$=0.60$ and KS $p=0.001$.
{\bf Middle:} the stream plane at $\Rv$ and the disc AM, with
$\la cos(\theta)\ra=0.45$, median$=0.43$ and KS $p=0.03$.
{\bf Right:} the AM at $\Rv$ and the disc AM, with
$\la cos(\theta)\ra=0.14$, median$=0.14$ and KS
$p=2\times10^{-4}$.
The AM at $\Rv$ is restricted to $V_{r}<-0.5\Vv$.
Note that $\hat{n}$, the normal to the SP, has no preferred orientation,
so the associated range of cosine values is $(0,1)$ rather than $(-1,1)$.
}

\label{fig:planes_cos_pdf}
\end{figure*}

Here we provide a preliminary study that focuses on the relative orientations 
of different relevant characteristic planes, 
as defined by the inflowing and accumulated gas component. 
On the halo scale, near $\Rv$, we compare the normal to the stream plane 
(which in many cases is closely related to the pancake plane) 
to the direction of the AM in a shell about $\Rv$.  
Both are then compared to the direction of the AM of the disc,
\adb{defined as the AM of the gas in a sphere of radius $0.1\Rv$,\footnote{The
disc AM direction largely coincides with the normal to the disc plane as
defined by best fit to the mass distribution.}}
and then to the AM in shells at different radii inside and outside the halo.
In this paper, the study is limited to the gas component, 
and to the $z=2.5$ snapshot,
thus not considering the time delay between the stream crossing of the
virial radius and its arrival at the disc.
Recall that the AM is computed about an origin that 
approximates the centre of mass of the disc, as described in \se{haloes}.

\subsection{Visual comparison of planes}
\label{sec:visual}

In \fig{aitoff_r1} we display for three haloes
Hammer-Aitoff maps of the AM at $\Rv$ and compare the orientations
of the three relevant planes: 
the stream plane and AM at $\Rv$, and the disc plane.  
The upper panels refers to the total AM at $\Rv$ and rotated to the 
corresponding frame.
The bottom panels refers to the AM component along the direction of the disc AM,
and rotated accordingly to the disc frame.
The color represents the fraction of the AM in the different streams at $\Rv$.

These figures highlight several non-trivial features.
First, they suggest that the three planes are not necessarily aligned with each 
other, indicating that any correlation among them must be weak. 
Second, we get the impression that there is a tendency for most of the AM 
at $\Rv$ to be carried by one, dominant stream. 
Third, the three planes seem to be intersecting in one line, 
which tends to roughly coincide with the dominant stream.
These features repeat both for the AM relative to the total AM at $\Rv$
and for its component along the disc AM.
We will quantify these findings and attempt to interpret them below.

\begin{figure}
 \centering
      \includegraphics[width=0.5\textwidth]{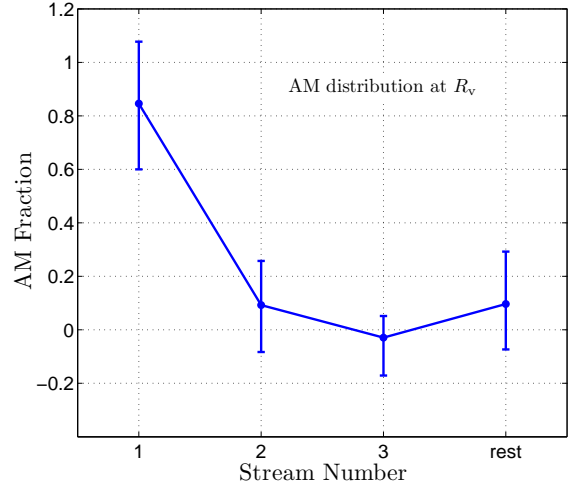}
\caption{
Distribution of AM among the streams at $\Rv$.
Shown are the mean and standard deviation of the relative contribution of the
N'th stream to the AM. Only cells with $V_r < -0.5\Vv$ are considered,
and the influx overdensity threshold is 2.
On average, one stream carries 84\% of the AM!
}
\label{fig:r1_am_vr_m2_dist}
\end{figure}

\begin{figure*}
\begin{minipage}[b]{0.5\linewidth}
 \centering
 \includegraphics[width=0.99\textwidth]{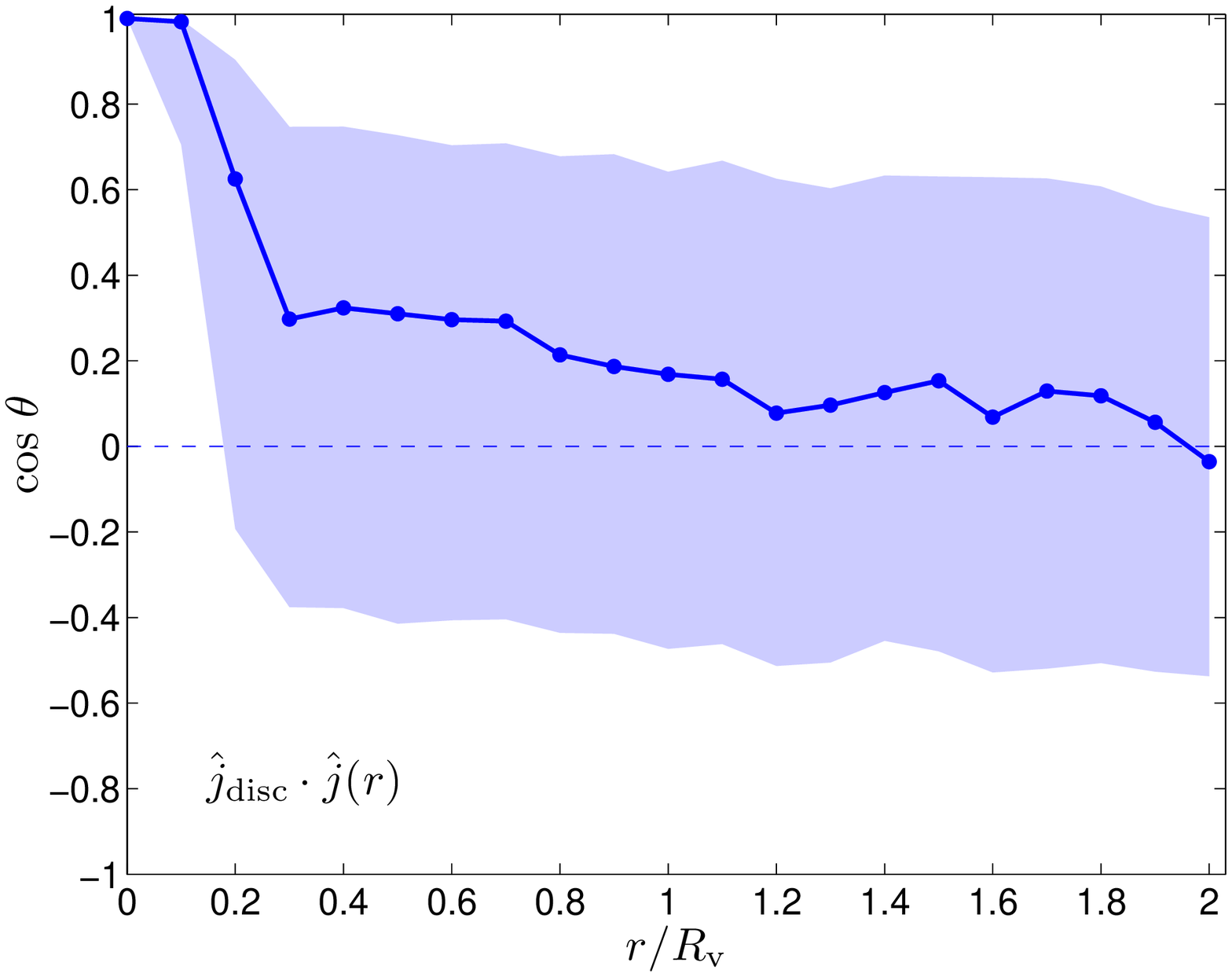}
\end{minipage}\hfill
\begin{minipage}[b]{0.5\linewidth}
 \centering
 \includegraphics[width=0.99\textwidth]{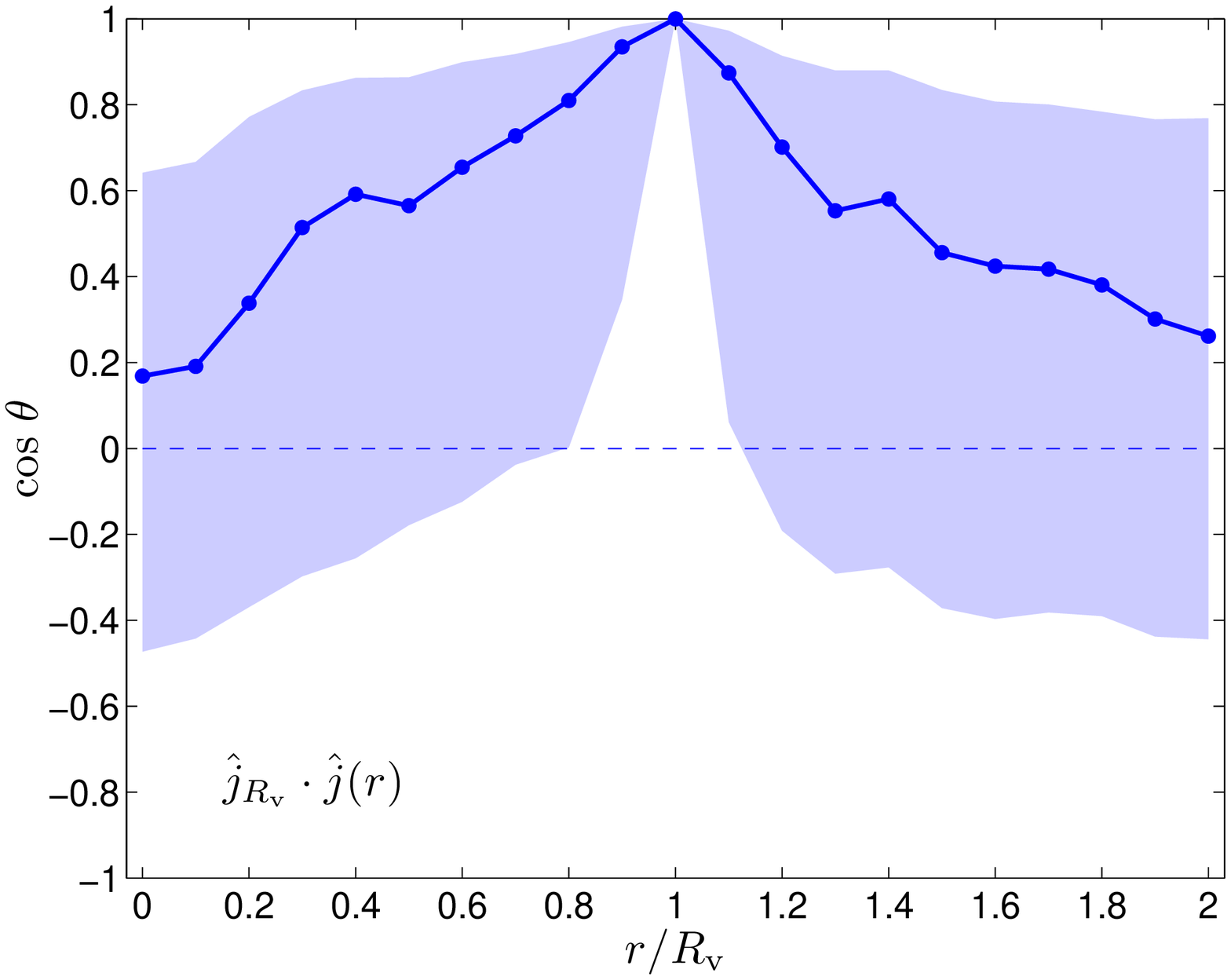}
\end{minipage}\hfill\\
\caption{
Alignment of AM in shells of different radii.
Shown is the median of the cosine of the angle between the AM of the disc
(left) or of a shell at $\Rv$ (right) and the AM in other shells at radius r
(and thickness $0.1\Rv$).
The shaded area represents the 1$\sigma$ scatter about the median
for the 336 haloes of the sample.
There is a tendency for alignment of AM from outside $\Rv$ down to $r \sim
0.3\Rv$, and independently in the immediate disc vicinity $r \leq 0.1\Rv$,
but there is an abrupt change of direction at $r = 0.1-0.3\Rv$.}
\label{fig:disc_am_cor}
\end{figure*}

\subsection{Stream plane versus AM at $\Rv$}

\Fig{planes_cos_pdf} (left) shows the probability distribution function (PDF) 
of the cosine of the angle between 
the stream plane and the AM at $\Rv$, limited to cells in which the inflow
velocity is larger than $0.5\Vv$ (a requirement that makes only a little
difference). 
The PDF shows a {\it weak tendency for alignment}, 
with a mean $\la cos(\theta)\ra =0.56$ and median$(cos(\theta))=0.60$
compared to $0.5$ for a random distribution.
The corresponding KS-test rejects the null hypothesis of a random angle
with a uniform distribution of cosines at a p level of only $0.001$.
This alignment being weak provides a hint that most of the AM is carried by one 
stream, consistent with the examples shown in \fig{aitoff_r1}.
\adb{A misalignment between the SP and the AM at $\Rv$ can occur if the
actual best-fit plane to the total influx slightly deviates
from the SP as defined here, where it is forced to include the galaxy center. 
This deviation is natural in the presence of transverse motions in the
streams, reflecting the flows from the sheets and voids into the streams.
These motions generate non-vanishing impact parameters that represent 
non-negligible AM, especially for the stream that dominates the AM.}

\subsection{AM Distribution among the streams}

\Fig{r1_am_vr_m2_dist} shows the distribution of AM among the streams at 
$\Rv$. The streams are ranked by their contribution to the total AM in the
shell, and the figure shows the mean and standard deviation of the relative
contribution of the N'th stream to the AM. 
For this figure, the influx overdensity threshold is 2, 
\adb{in order to include a significant fraction of the stream influx},
and only inflowing gas with $V_r < -0.5\Vv$ is considered,
\adb{in order to focus on the gas that will certainly reach the disc vicinity. 
The specific choice of these criteria for identifying the streams
do not affect the distribution of AM among the streams.}
We find that in almost all galaxies there is {\it one dominant stream\,} that
carries on average 84\% of the AM.
This 
\adb{helps explaining}
why the AM vector is not necessarily aligned with the stream
plane at the same radius.
In 70\% of the galaxies the stream that dominates the AM is also
the stream that carries the largest influx of mass, but in some cases
a large impact parameter makes a stream of lower influx carry most of the AM. 


\subsection{The stream plane versus the disc}

\Fig{planes_cos_pdf} (middle) shows the PDF of the cosine of the angle between 
the stream plane and the disc AM 
in the whole sample of 336 simulated galaxies.
It shows a weak tendency of the disc AM plane
to be perpendicular to the streams plane, with $\la cos(\theta)\ra=0.45$ 
and a median of $0.43$ compared to $0.5$ for a random angle, and a KS 
$p=0.03$. 
About 60\% of the haloes have $cos\theta<0.5$. 
A correlation of this nature is consistent with tidal-torque theory, 
which predicts a tendency for perpendicularity between the AM vector of the 
galaxy and the intermediate axis of the tidal field,
as demonstrated for dark-matter haloes in cosmological N-body simulations
\citep[e.g.][]{porciani02a,porciani02b,navarro04,aragon07}.
However, our main finding is that this correlation becomes negligibly small
for galaxies at the studied highly non-linear stage of evolution.
In comparison, there were marginal observational detections of alignments
between isolated local disc galaxies and the intermediate axis of the
tidal tensor \citep{dekel85,navarro04}, 
as well as failures to detect such an alignment for the massive galaxies 
at high redshift as studied in hydrodynamical simulations \citep{hahn10}.

\subsection{AM at $\Rv$ versus the disc: AM exchange}

\Fig{planes_cos_pdf} (right) shows the PDF of the cosine of the angle between
the AM at $\Rv$ ($V_r\!<\!-0.5\Vv$) and the disc AM.
The tendency for alignment is stronger than in the other cases, 
with a KS $p=2\times10^{-4}$ compared to a random distribution,
but it is still surprisingly weak, with $\la cos(\theta)\ra=0.14$ and a 
median of $0.14$ compared to zero for the case of no correlation.
The excess seen at $\cos\theta >0.6$ and the deficiency below $-0.4$
each involves only $\sim 10\%$ of the galaxies.
One may suspect that the alignment may be weakened by us not considering
the time delay between $\Rv$ crossing and
reaching the disc, which is on the order of $500 \Myr$.
To the extent that the orientation of the AM at $\Rv$ and the disc plane
are not varying drastically over this timescale, we learn that the AM is
typically not conserved all the way to the disc.

\begin{figure}
\includegraphics[width=0.49\textwidth]{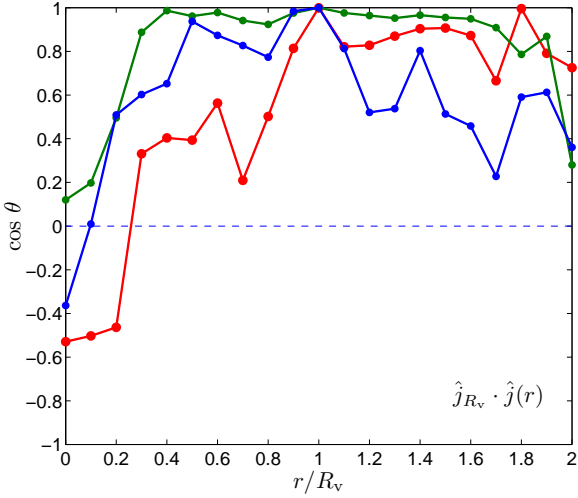}
\caption{
Three realizations from the distribution shown in \Fig{disc_am_cor} (right) 
of the alignment between the AM at $\Rv$ and in shells of other radii,
displaying different behaviours.
The green line (galaxy 250) shows a strong alignment from 
$r=1.9\Rv$ down to $0.3\Rv$, followed by a drop in $\cos(\theta)$
inside this radius toward zero.
The red line (66) shows an alignment outside $\Rv$ and a change in direction
immediately inside $\Rv$ leading to a slight anti-alignment inside $0.25\Rv$.
The blue line (219) is an intermediate case, qualitatively similar to the 
median shown in \Fig{disc_am_cor}.
}
\label{fig:disc_am_cor_ex}
\end{figure}

\begin{figure}
  \centering
\includegraphics[width=0.45\textwidth]{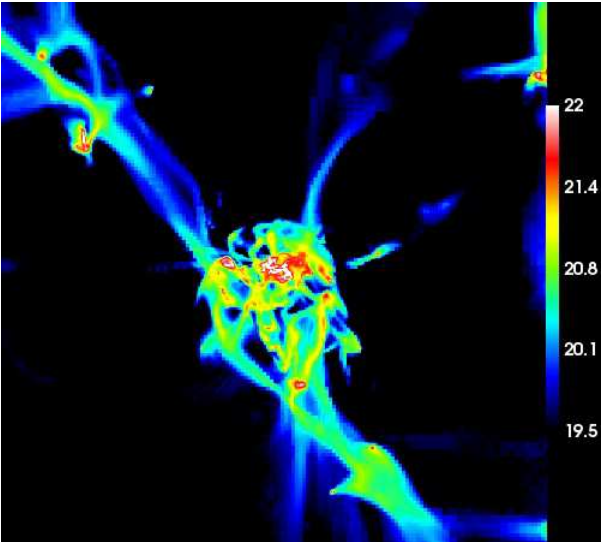}
\caption{Surface density of cold gas in a galaxy simulated with 70-pc
resolution \citep{cdb10}. Beyond the coherent incoming streams coming from
more than $100\kpc$ away, and the inner disc of radius $\sim 6\kpc$ (white),
there is a "messy" region of radius $\sim 20 \kpc$ in which the perturbed
distribution of matter results in strong torques and AM exchange --- the AM
sphere.
}
\label{fig:messy}
\end{figure}

\Fig{disc_am_cor} addresses the alignment between the AM in shells of different
radii, and it describes our most interesting result concerning AM conservation
in disk formation. 
It shows the median of the cosine of the angles, once with respect to
the disc AM (left) and once with respect to the shell at $\Rv$. 
The shells are of thickness $0.1\Rv$.
We see a tendency for alignment of AM from outside $\Rv$ down to $r\sim0.3\Rv$,
and in the disc and its immediate vicinity, $r \leq 0.15\Rv$, but an abrupt
change of AM direction at $r = 0.2-0.3\Rv$. 
This marks the inner sphere within which there is 
a significant exchange of AM, and at the centre of which the actual
disc orientation is almost arbitrary compared to the AM at $\Rv$. 
We term this ``the AM sphere". 

\Fig{disc_am_cor_ex} shows three concrete realizations from the sample that
makes the average shown in the right panel of \fig{disc_am_cor}.
It shows one case where the AM direction is rather
constant from well outside the halo down to the boundary of the 
AM sphere at $\sim 0.3\Rv$, inside which the AM direction significantly 
deviates from the AM at $\Rv$.
It also shows an opposite case where the AM inside most of the virial sphere
is far from being aligned with the AM at and outside $\Rv$.
The third case is rather typical, similar to the average shown in
\fig{disc_am_cor}.

\subsection{The AM Interaction Sphere}
\label{sec:messy}

One should not be totally surprised 
\adb{by}
the weak alignment between the AM 
on the halo scale and in the disc vicinity, given what we already have a 
preliminary notion about how the in-streaming gas joins the disc.
\Fig{messy} shows the surface density of cold gas in a typical galaxy 
simulated at 70-pc resolution and described in \citet{cdb10}.
It shows 
\adb{at large radii}
coherent incoming streams that emerge from outside the virial radius
and penetrate 
\adb{toward}
the inner halo, to 
\adb{eventually}
end up as an inner disc of radius 
$\sim 6\kpc$. 
\adb{However, the streams break up by collisions, shocks and various 
instabilities before they reach the disc.}
Surrounding the disc there is a ``messy" region of radius 
$\sim 20 \kpc$, which consists of stream fragments and clumps moving in and 
out in a complex pattern and kinematics.
This highly perturbed distribution of matter,
\adb{including the disc itself,}
is naturally associated
with strong torques and significant AM exchange between the different gas
components and possibly between gas, stars and dark matter.

\subsection{Intersection of planes and the dominant stream}

\begin{figure*}
\begin{minipage}[b]{0.25\linewidth}
 \centering
 \includegraphics[width=1.02\textwidth]{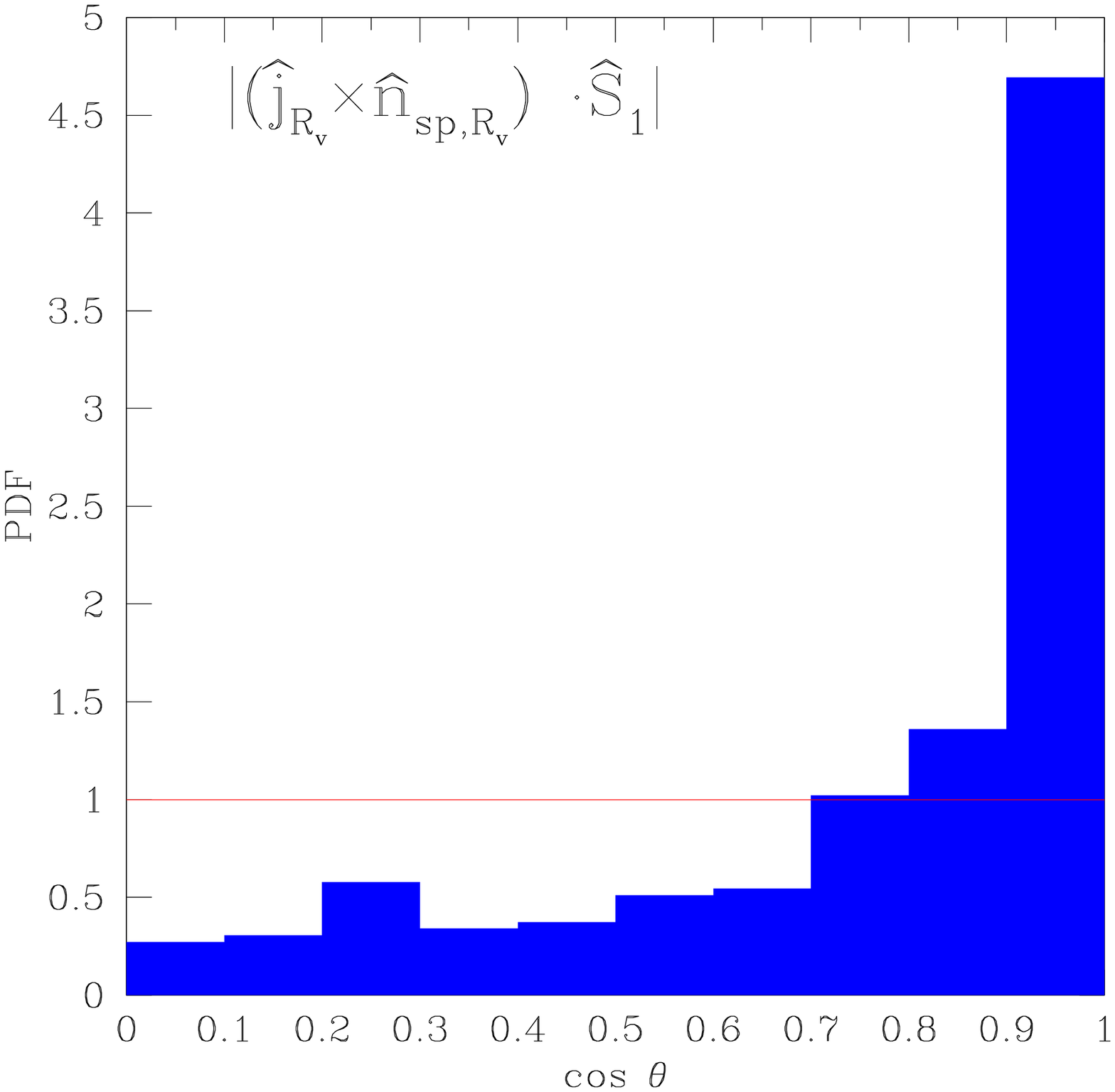}
\end{minipage}\hfill
\begin{minipage}[b]{0.25\linewidth}
 \centering
 \includegraphics[width=1.02\textwidth]{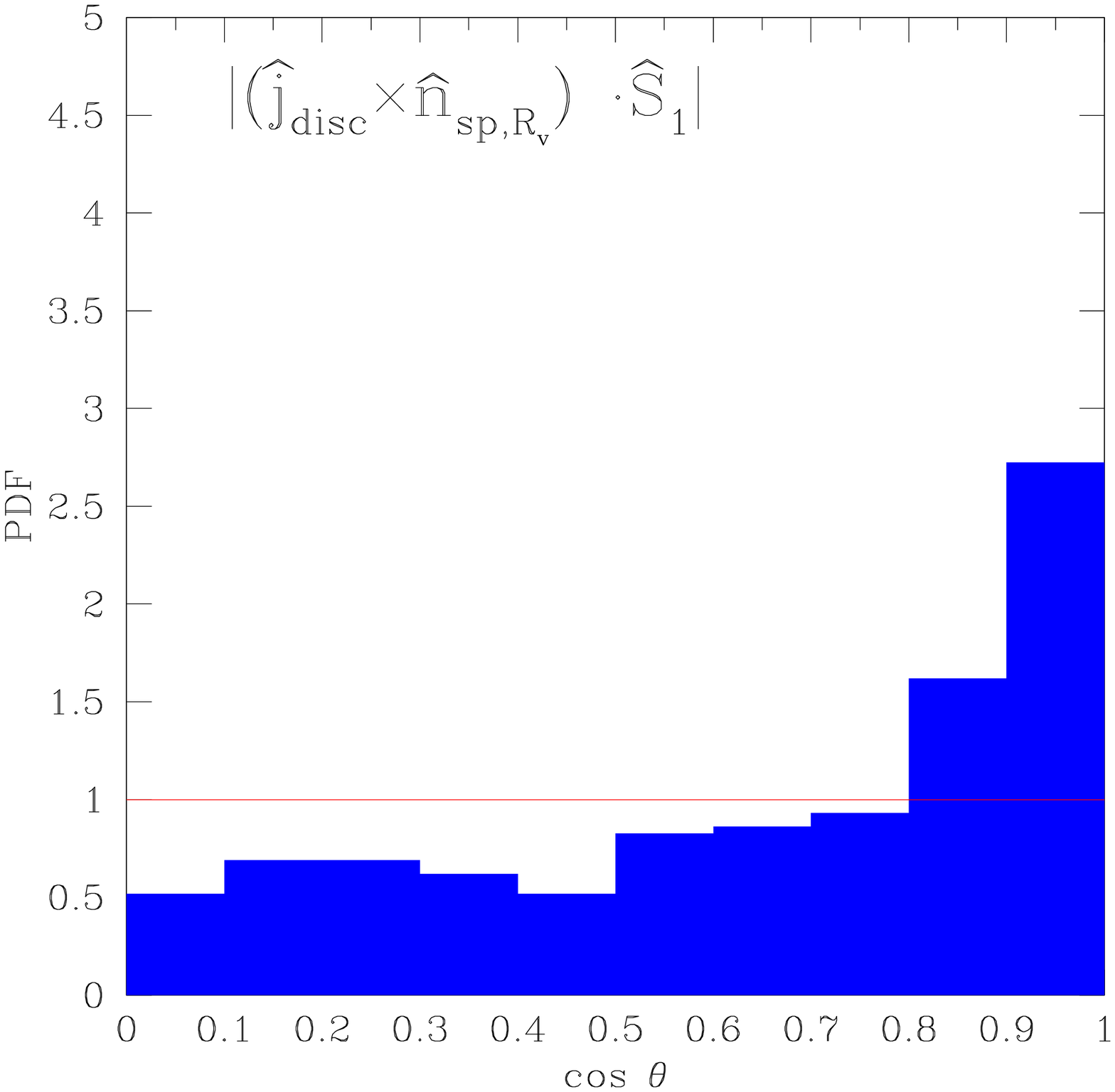}
\end{minipage}\hfill
\begin{minipage}[b]{0.25\linewidth}
 \centering
 \includegraphics[width=1.02\textwidth]{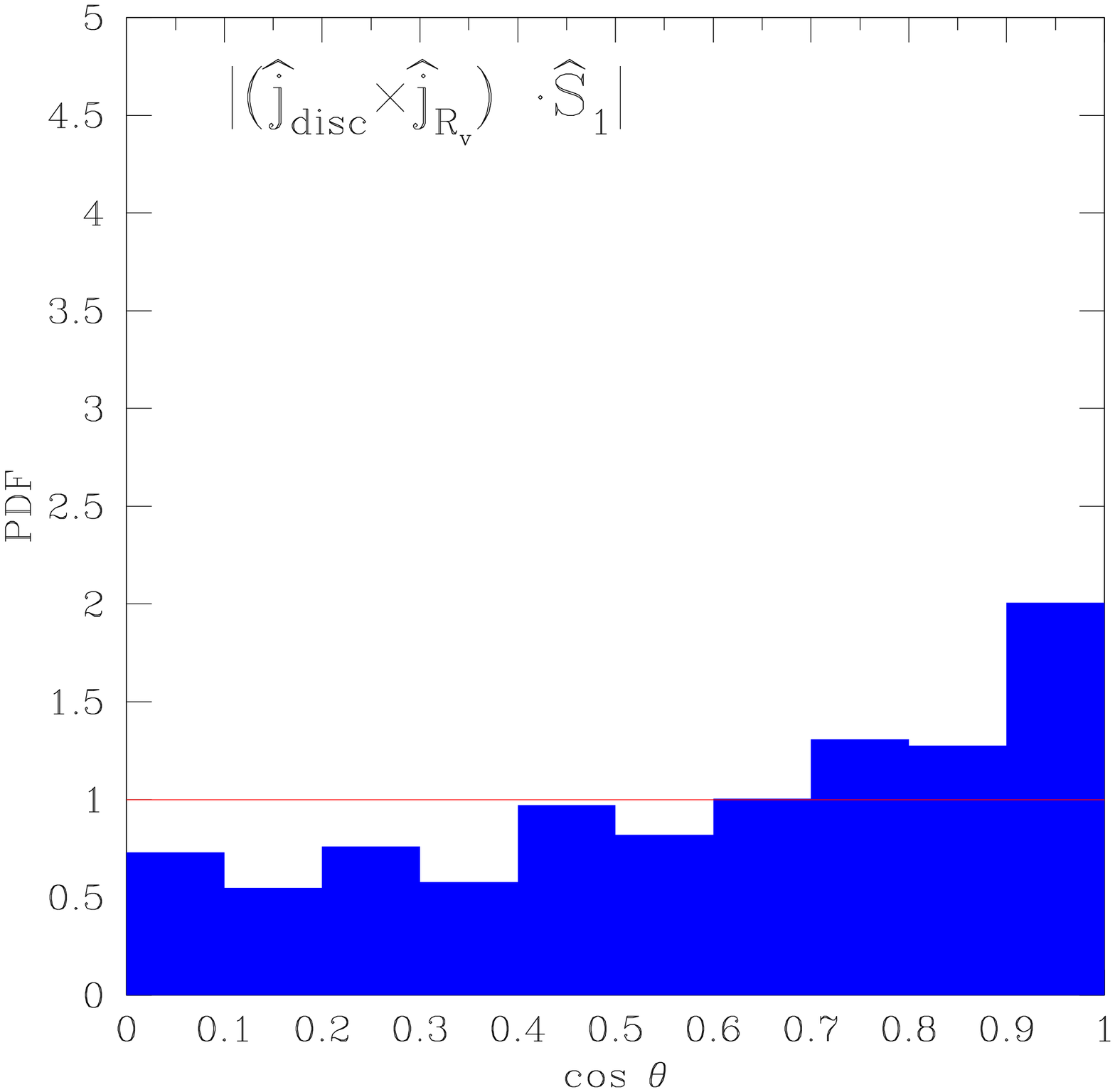}
\end{minipage}\hfill
\begin{minipage}[b]{0.25\linewidth}
 \centering
 \includegraphics[width=1.02\textwidth]{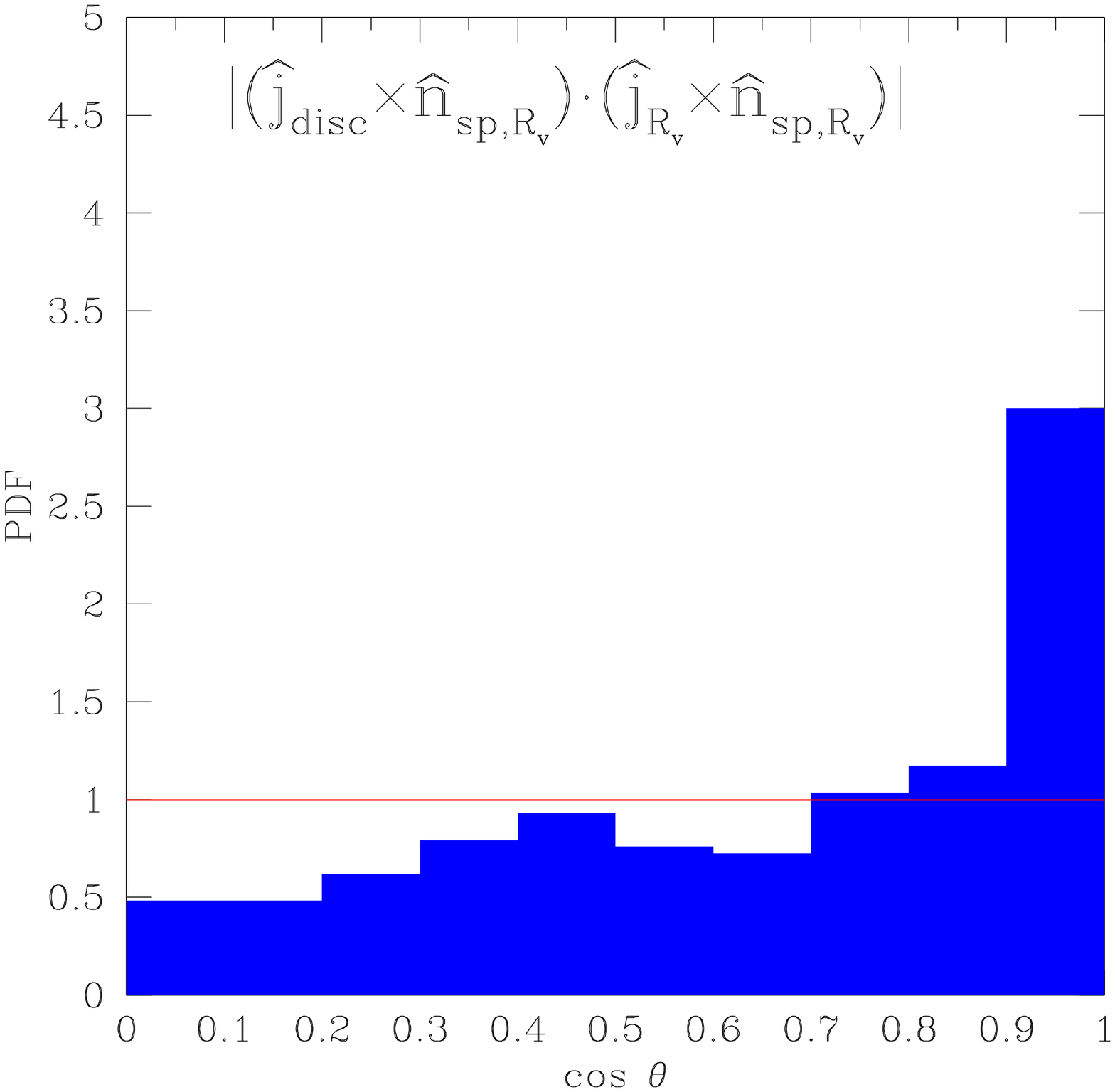}
\end{minipage}\hfill\\
\caption{
Alignment between the intersection of planes and the dominant stream:
the AM at $\Rv$ ($\hat j_{\Rv}$),
the normal to the stream plane ($\hat n_{\rm sp}$),
the disc AM ($\hat j_{\rm disc}$),
and the vector along the dominant stream at $\Rv$ ($\hat S_1$).
The means are: 0.76, 0.65, 0.60, 0.71.
The medians are: 0.88, 0.73, 0.66, 0.71.
The KS p values are smaller than unity by many orders of magnitude in all
cases.
}
\label{fig:cross}
\end{figure*}

Back to what we learned from a visual inspection of \fig{aitoff_r1}
regarding the apparent tendency of the three planes to 
intersect in one line, which is the line defined by the dominant stream
in terms of influx of mass and AM.
\Fig{cross} shows the PDFs of cosine of angles between pairs of lines,
each corresponding to the intersection of two of the planes or by the dominant
stream $\hat S_1$.

The strongest alignment is between the dominant stream and the line of
intersection between the stream plane and the AM at $\Rv$, 
namely, $\hat j_{\Rv} \times \hat n_{{\rm sp},\Rv}$, with a median
at $\cos\theta=0.88$. This is fully consistent with the fact that a single
stream indeed carries most of the AM at $\Rv$, and it explains why the
stream plane and the AM at $\Rv$ are not necessarily aligned with each other.

The dominant stream also tends to be aligned with the lines of intersection
between the disc and the two planes at $\Rv$, the stream plane and the AM at
$\Rv$, namely, $\hat j_{\rm disc} \times \hat n_{{\rm sp},\Rv}$ 
and $\hat j_{\rm disc} \times \hat j_{\Rv}$,
but these alignment are weaker, with medians at $\cos\theta = 0.73$
and $0.66$ respectively. 

Finally, the three planes, despite their very weak alignment with each other,
do tend to intersect along one line, as seen by the PDF of 
the absolute value of $(\hat j_{\rm disc}\times\hat n_{\rm sp})\cdot
(\hat j_{\rm \Rv}\times\hat n_{\rm sp})$,
for which the median is $\cos\theta = 0.71$.

\smallskip

The emerging picture is that typically {\it one stream plays the major role in 
bringing both the mass and the angular momentum into the halo}.
This stream determines the AM in the whole inner $0.3\Rv$ of the halo, 
the AM sphere.
However, the AM in this sphere is only partially reflected in the direction 
of the disc AM due to {\it significant AM exchange inside the AM sphere}. 

\section{Discussion: on the origin of the web structure about a node}
\label{sec:theory}

Our analysis of the simulations reveal three robust features of the cosmic web
in the vicinity of a high-sigma node, namely, the dominance of a single
sheet, the dominance of one filament in it, and the preferential
contribution from three main filaments embedded in that sheet.
In principle, these features should be understood either in terms of the
statistics of the initial Gaussian random fluctuation field,
or considering the later non-linear evolution involving motions and mergers of
structures, or both.
This distinction between early and late evolution at a given scale
can be replaced by large-scale smoothed linear structure versus small-scale
non-linear structure at a given time.
While these are open theoretical challenges beyond the scope of the present
paper, we mention here preliminary ideas concerning these issues.

As mentioned in \se{intro}, based on the Zel'dovich approximation,
one can identify the four basic structures of voids, sheets, filaments
and haloes with the four possible
signatures of the eigenvalues of the local deformation tensor.
The spatial extent of each sheet or filament is determined by the coherence
length of the relevant eigenvectors.
Since a sheet requires a coherence of only one eigenvalue, while a filament
requires the simultaneous coherence of two, which is less probable,
one can expect the sheets to be in general more extended than the filaments,
favoring a configuration of filaments that are embedded in larger sheets,
as detected in the simulations.
The spatial coherence of the eigenvectors is expected to be especially extended
near a node, which can be treated as a high-sigma density peak in the Gaussian
random field.  The coherence length can in principle
be estimated by computing the conditional probability for the value
of a given eigenvalue at a distance $r$ from a density peak, 
in which the density contrast
$\delta\rho/\rho \propto (\lambda_1+\lambda_2+\lambda_3)$ is given and is high.
The average value of the eigenvalue at $r$ is provided by the corresponding
conditional two-point correlation function. Most important, the variance
about this conditional average is limited by the general variance over space,
so the coherence length is expected to be larger about a higher density peak
\citep{dekel81,bbks}. 
We thus expect long filaments embedded in even more extended sheets
to form at an early stage about the massive galaxies that form at
the cosmic-web nodes.

The tendency toward a minimum number of filaments and sheets in the vicinity of
a node could be understood in terms of the continuity of the eigenvectors
of the deformation tensor within the smoothing length $\rs$ of the fluctuation
field.
Within one smoothing length from the node, the eigenvalues at the different
points must converge to the vicinity of one set of values. 
This allows only one dominant
filament embedded in one dominant sheet throughout the smoothing volume.
For example, two different filaments through a node, with an opening angle
of $90^\circ$, say, would require that the eigenvector with the smallest
eigenvalue of one filament coincides with one of the eigenvectors with
the large eigenvalues of the other filament, which would violate the coherence
of the eigenvectors within the smoothing volume.
This implies that two filaments with a large opening angle
between them far away from the node must merge into a single filament
within the $\rs$ vicinity of the node (which
may correspond in our analysis to two streams
inflowing toward the node in opposite directions along the same line).
The continuity implies that the filaments merge smoothly in a bifurcation
point at a distance $\sim \rs$ from the node, such that they form
a swallow-tail structure, like a splitting fork in a road,
similar to what is predicted in Fig. 10 of \citet{arnold} and hinted in 
\fig{bird} above.
Analogous considerations apply to the presence of one dominant sheet.
The relevant smoothing scale in our study is the halo virial radius
about the node, which stretches to a Lagrangian region of
a few virial radii (a mean density contrast of 180 within
$\Rv$ translates to a comoving radius of $5.6\Rv$).
We therefore expect the sheet about a node to extend to a few virial radii,
as detected in the simulations.

A somewhat different quantitative approach to the properties of the cosmic web
is based on the ``skeleton" of critical lines (filaments) connecting
critical points (nodes), mastered by \citet{novikov06},
\citet{pogosyan09} and \citet{pichon10}.
It makes use of the Hessian $H$, the tensor consisting of
the second spatial partial derivatives of the density field,
whose local eigenvalues are denoted $h_1 \geq h_2 \geq h_3$.
The primary skeleton is defined as the set of points where $h_1+h_2 \leq 0$
and where the gradient of the density
field is an eigenvector of the Hessian, and corresponds to the largest
eigenvalue, $H\cdot\nabla\rho=h_1\nabla\rho$.
In this formalism, the filaments connect the density maxima at the nodes
and the saddle points mid-way between maxima.  The eigenvector corresponding to
$h_1$ defines the preferred direction for the dominant filament
within the $\rs$ vicinity of the maximum.
Outside $\rs$, one expects the single filament to bifurcate into two where
$h_1=h_2$, i.e., where the two directions are locally equivalent.
This creates three filaments in a plane, as detected.
The sequence of bifurcations as one receded from the node indicates
a shift from two filaments along a single line inside the virial radius,
to three filaments in a plane near and outside the virial radius, and to
a larger number of filaments in more sheets further away from the nodes,
until these filaments construct the three-dimensional web on scales
much larger than $\rs$.

Being fed by three dominant filaments at the virial scale
is not an obvious property of the nodes in a general three-dimensional network.
For example, in a symmetric cubic lattice each node is fed by six filaments
along three lines.
On the other hand, with only two filaments per node, one cannot construct
a three-dimensional web, not even a two-dimensional network, so three is the
minimum average number of filaments per node.
Once we found that the filaments near a node tend to be confined to one plane,
and obtained hints concerning the possible origin of this dominant plane,
the question reduces to the appearance of three filaments in the effective
two-dimensional space about the node.
This issue has been addressed in the early mathematical analysis of caustics
that bound filaments and sheets following the quasi-linear description
of structure development a la Zel'dovich.
\citet{arnold,shandarin89} showed in this analysis that a node of three 
filaments is the natural configuration in two-dimensional space.
They also argue that the swallow-tail configuration of these three filaments
is common.
This theory of caustics is yet to be properly related to the structure in a
three-dimensional Gaussian random field.

In two dimensions, a symmetric configuration with three filaments per node
is the hexagonal grid, or the honeycomb, as opposed to the cubic grid with
four filaments per node.
The empty spaces in a honeycomb encompass bigger spherical voids than in
a cubic grid, and since the underdense voids are dynamically driven
to become spherical by the gravitational forces, this dynamical effect
may drive the cosmic web into a honeycomb configuration.
Another possibly relevant property of the hexagonal grid, known as the
honeycomb theorem \citep{hales01},
says that the total length of filaments in a honeycomb is the minimum possible
length among all the partitions of the plane into regions of equal area.
However, the direct relevance of this to the cosmic web is yet to be
addressed.


\section{Conclusion}
\label{sec:conc}

Our results shed new light on how massive galaxies at high redshift
acquire their mass and angular momentum.
It has been known for a while, based on cosmological simulations, that
the baryons flow in along narrow streams that follow the dark-matter
filaments of the cosmic web toward the high-density peaks at the nodes where
they intersect 
\citep{bd03,keres05,db06,ocvirk08,dekel09}.
These streams consist of cold gas and a spectrum
of merging galaxies.
We now find that at the few virial radii vicinity of the galaxy, 
the streams tend to be confined to a stream plane, 
and embedded in a flat pancake that carries $\sim 20\%$
of the influx.
There are on average three significant streams, of which one typically carries
more than half the mass inflow.  

This structure of filaments that are embedded in an extended sheet is unique
to the neighborhood of massive galaxies at high redshift, where they reside
in the high-sigma density peaks that are associated with
the nodes of the cosmic web.  On larger scales, the 
filaments are the intersections of coherent sheets that are tilted relative 
to each other and together encompass big voids.
The transition from a three-dimensional web to a planar distribution
of streams on the virial scales near the nodes introduces
a non-trivial theoretical
challenge. So is the tendency to have three major streams, of which one
dominates. We mention very crude hints for the origin of these phenomena
in section \se{theory}.

Small transverse velocities of the streams at large distances induce  
non-zero impact parameters of the streams relative to the galaxy centre, 
and the associated angular momentum is 
transported with the streams into the galaxy 
\citep[see also][]{pichon11,kimm11}.
The fact that at later times the gas originates from larger distances indicates
that it carries larger specific AM, which gives rise to disk growth inside out.
The major stream typically carries $\sim 80\%$ of the AM near $\Rv$.
This dominant stream and the galaxy centre define the AM plane in the outer
halo and outside it, which does not necessarily coincide with the stream plane.
The AM direction is preserved as the stream penetrates the outer halo
and till it reaches $r\sim 0.3\Rv$.
Inside the sphere of $0.3\Rv$, the coherent streams shock, break and interact 
with the disc, so the mass distribution and kinematics become asymmetric and
complex. 
This leads to significant AM exchange between the different
components due to strong torques. 
The disc orientation at the centre of this AM sphere turns out to be
 quite arbitrary.

Much of the analytic and semi-analytic modeling of disc formation 
is based on the very useful simplifying
assumption that the gas conserves its AM as it flows in through the halo,
implying that the disk radius scales with the virial radius 
times a constant spin parameter $\sim 0.05$
\citep{fall_efstathiou80,bullock01,mo_mao_white}.
Our results here indicate that this assumption is invalid as far as the
direction of the disc AM is concerned. However, the jury is still out 
on the extent to which the amplitude of the disc AM approximates 
the AM of the inflowing mass, and its implications on the disc size and
inner structure. 
This is work in progress.
We note that the significant misalignment may be associated with a change
of only a factor of two in the spin amplitude.
This is similar to the results concerning the validity of the tidal-torque
theory in predicting the direction and amplitude of the halo spin
\citep{porciani02a}.

Our evaluation of the limited artificial tendency of the discs to align 
with the simulation grid (\se{axes})
indicates that our conclusions are not biased by this numerical effect.
A much stronger artificial effect is required in order to hide a strong 
alignment between the disc and the AM at $\Rv$.
Indeed, similar studies using zoom-in cosmological simulations with a 
resolution more than ten times better reveal similar results 
\citep{hahn11}.

\adb{
The direct relevance of the results obtained here for massive discs at high 
redshift to low-redshift discs should be considered with caution, 
as the latter probably develop under slow, wide-angle accretion rather then 
by the intense, narrow, high-redshift streams \citep{db06}. 
We expect many of the massive discs analyzed here at $z=2.5$
to evolve into early-type galaxies at low redshift \citep{dsc09,cdb10}, 
while today's discs arise from smaller progenitors, typically with quiet 
merger histories after $z \sim 1$ \citep{martig12}, 
and in which feedback slows down star formation and allows the late formation 
and maintenance of thin discs \citep{governato10,guedes11}.
}

\section*{Acknowledgments}

We acknowledge stimulating discussions with S. Colombi, J. Devriendt,
C. Pichon, and N. Schenkler. 
This work was partially supported by ISF grant 6/08, 
by GIF grant G-1052-104.7/2009, by a DIP grant, and by NSF grant AST-1010033.
The Mare Nostrum simulation was run on the
Barcelona Centro Nacional de Supercomputacion as 
part of the Horizon collaboration.

\bibliographystyle{mn2e}
\bibliography{paper31}

\bsp


\begin{figure}
\includegraphics[width=0.49\textwidth]{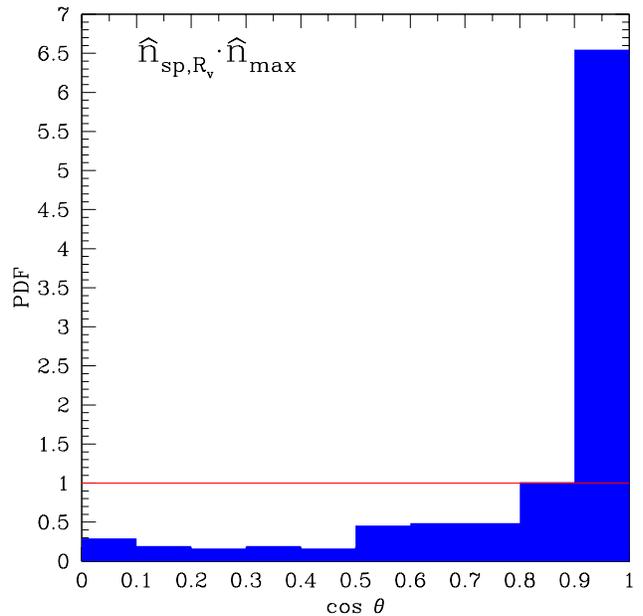}
\caption{
PDF of the cosine of the angle between the stream plane as defined
in \se{fit} and the maximum-influx plane.
We see a strong alignment, with a median of $\cos\theta=0.96$.}
\label{fig:strip}
\end{figure}

\begin{figure*}
\begin{minipage}[b]{0.33\linewidth}
      \centering
      \includegraphics[width=1.1\textwidth]{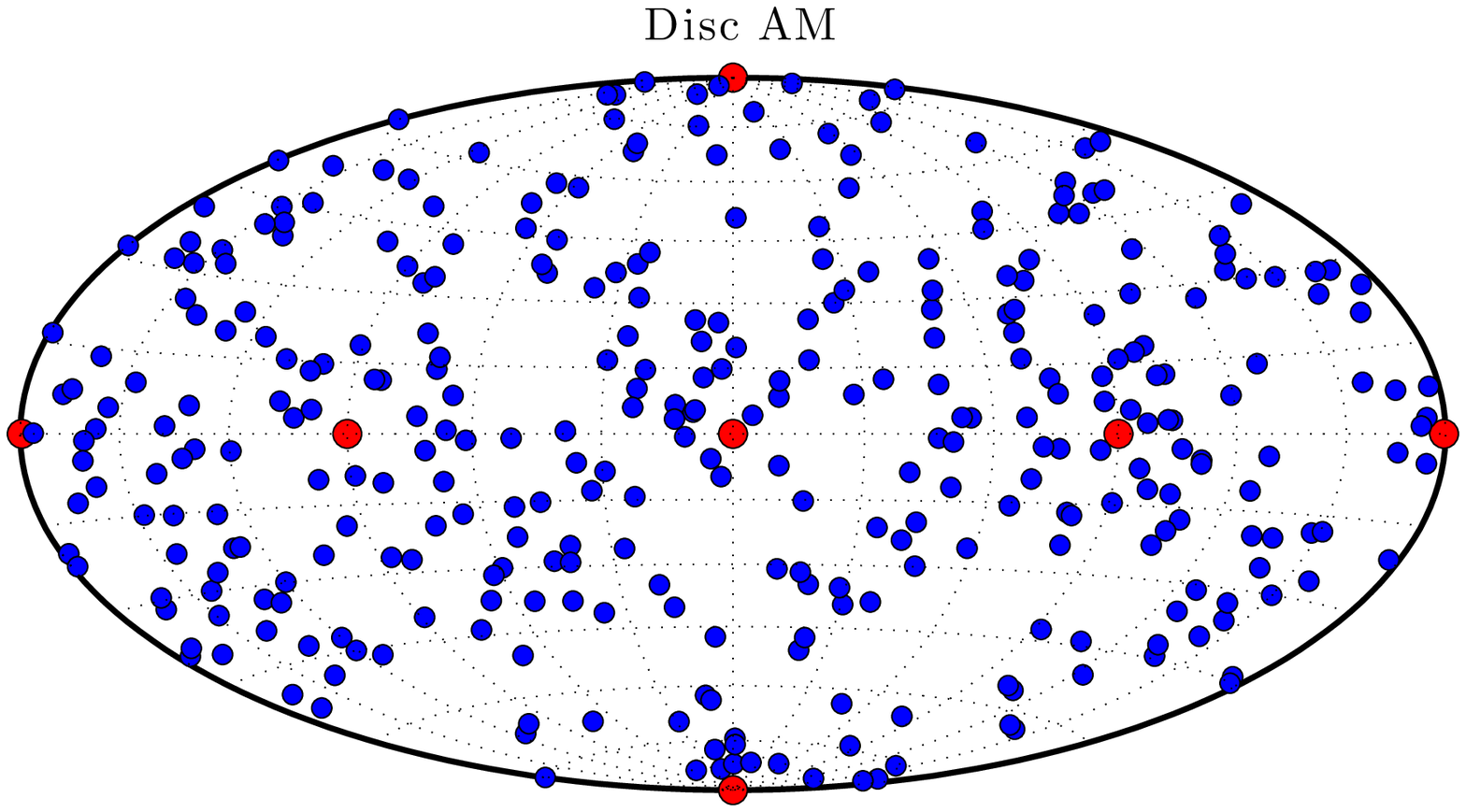}
  \end{minipage}\hfill
\begin{minipage}[b]{0.33\linewidth}
      \centering
      \includegraphics[width=1.1\textwidth]{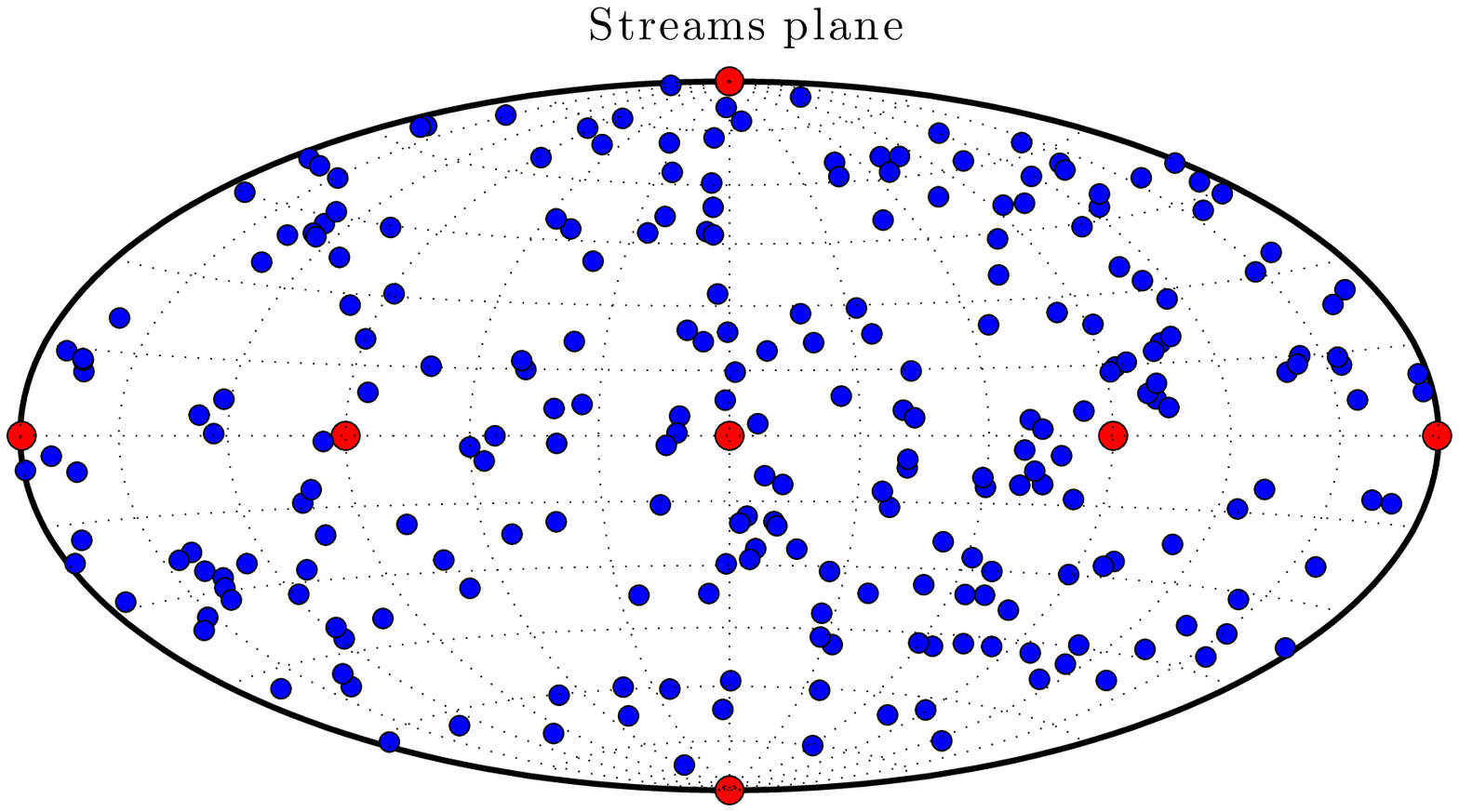}
  \end{minipage}\hfill
  \begin{minipage}[b]{0.33\linewidth}
      \centering
      \includegraphics[width=1.1\textwidth]{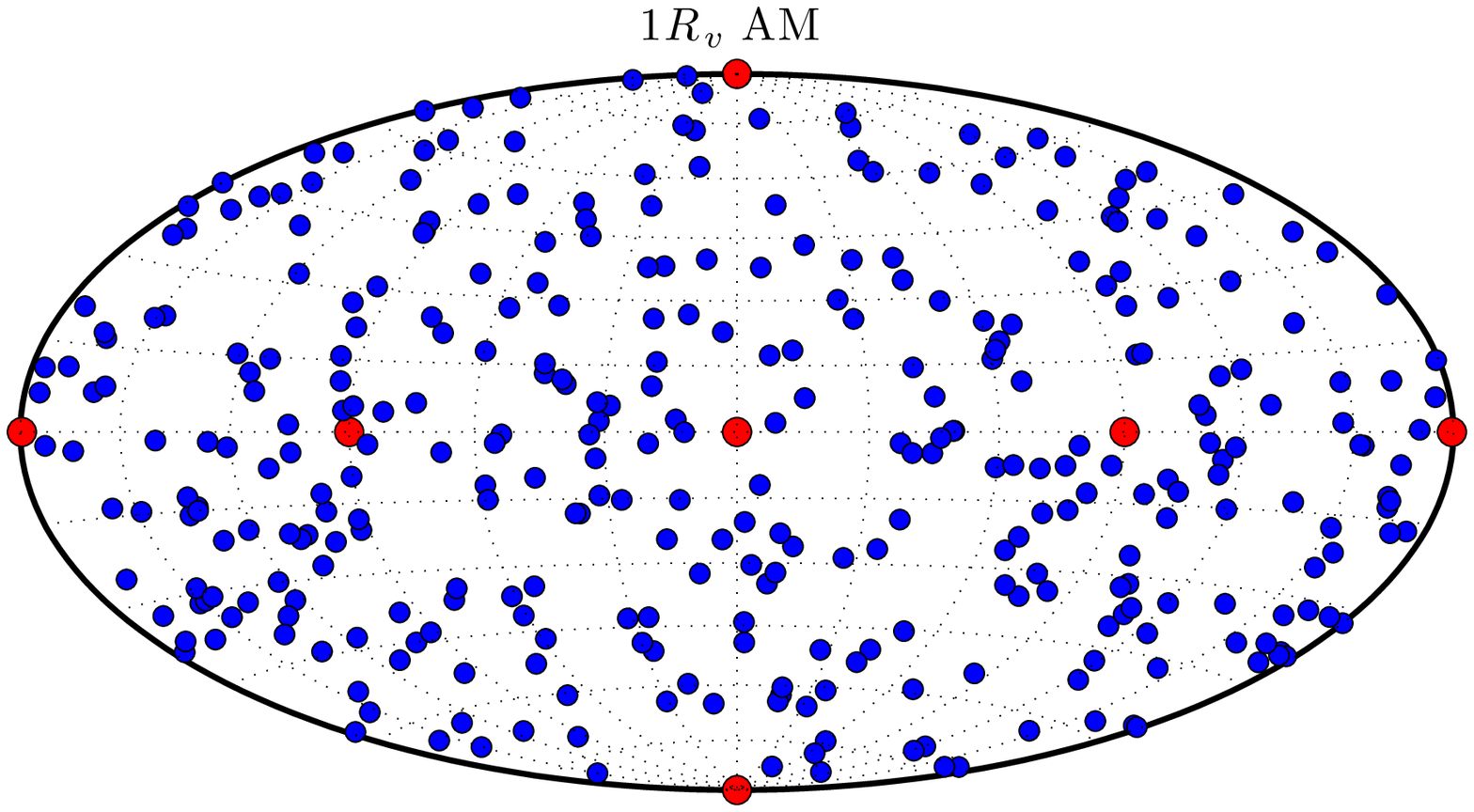}
  \end{minipage}\hfill\\
\caption{
Numerical alignment.
Hammer-Aitoff projection showing for each galaxy the direction of the normal
to the plane in question (blue dots).
The red dots represent the 6 directions of the simulation grid axes.
{\bf Left:} disc plane.
{\bf Middle:} stream plane at $\Rv$.
{\bf Right:} angular momentum at $\Rv$.
No obvious clustering of the blue dots about the red dots is noticeable,
indicating that the numerical alignment with the simulation grid is weak.
}
\label{fig:sim_bias_aitoff}
\end{figure*}

\begin{figure*}
\begin{minipage}[b]{0.33\linewidth}
      \centering
      \includegraphics[width=1\textwidth]{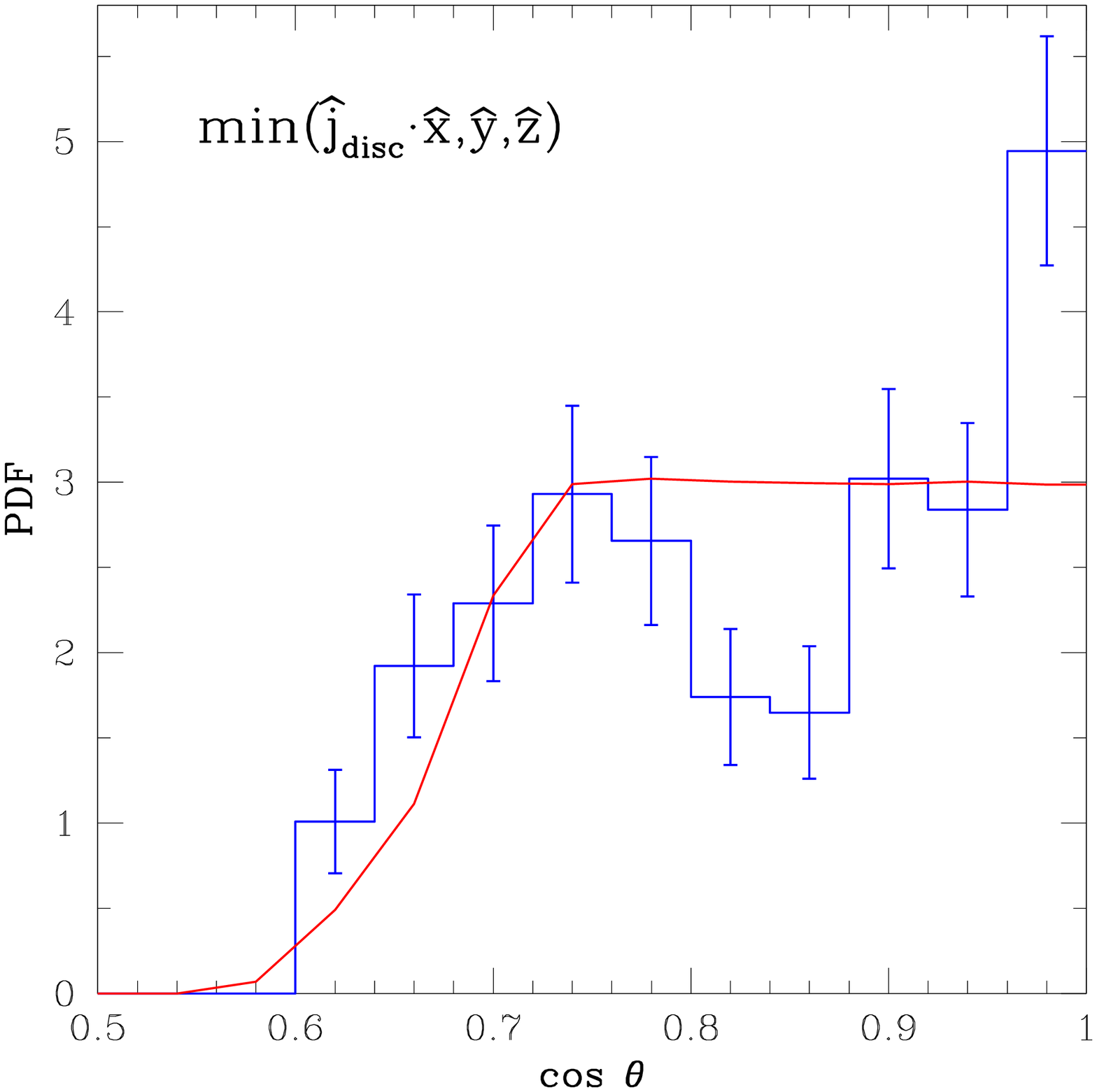}
  \end{minipage}\hfill
\begin{minipage}[b]{0.33\linewidth}
      \centering
      \includegraphics[width=1\textwidth]{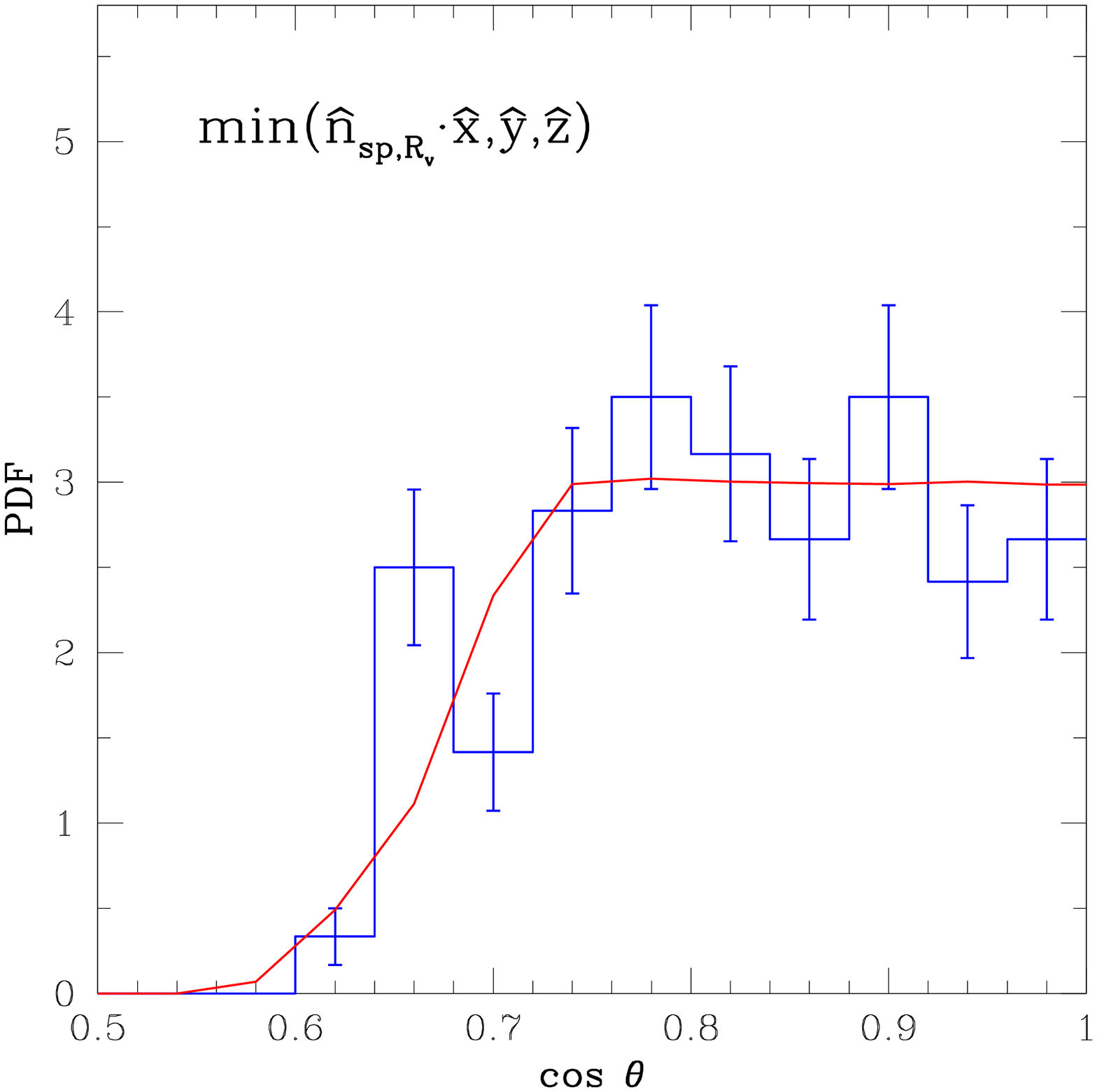}
  \end{minipage}\hfill
  \begin{minipage}[b]{0.33\linewidth}
      \centering
      \includegraphics[width=1\textwidth]{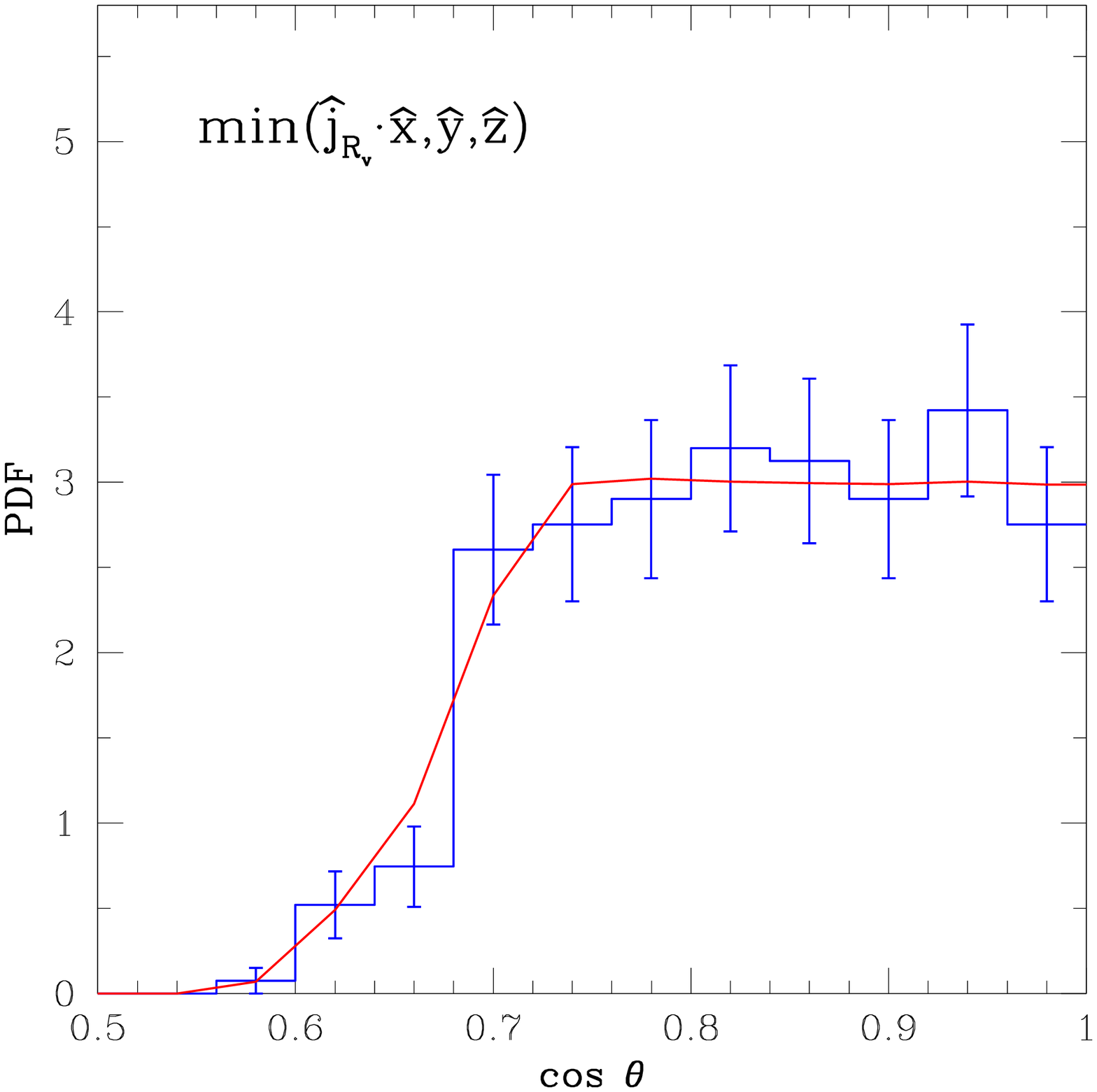}
  \end{minipage}\hfill\\
\caption{Numerical alignment. The distribution of the cosine of the angle
between the normal to the plane in question
and the closest simulation grid axis (blue).
It is compared to the corresponding distribution for random plane
orientations (red).
{\bf Left:} disc plane.
{\bf Middle:} stream plane at $\Rv$.
{\bf Right:} angular momentum at $\Rv$.
The disc shows a weak numerical alignment signal involving $\sim 20\%$ of the
discs at the level of $\Delta \cos \theta \sim 0.08$.
}
\label{fig:sim_bias_pdf}
\end{figure*}

\begin{figure*}
      \centering
      \includegraphics[width=1\textwidth]{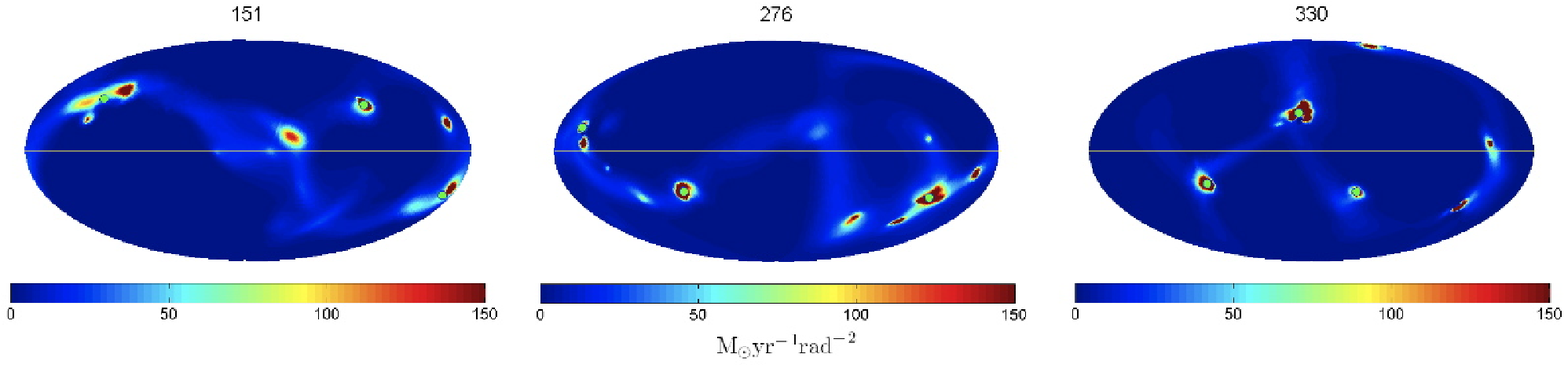}
\caption{Poor stream planes.
Hammer-Aitoff projection for 3 haloes, similar to \Fig{planarity_ex},
showing the three worst cases for a stream plane.}
\label{fig:bad_planarity}
\end{figure*}

\appendix

\section{The Mare Nostrum Simulation}
\label{sec:MN}

The Horizon Mare Nostrum cosmological simulation follows the evolution of a
cubic box of side $50\hmpc$ (comoving), containing $1024^3$ dark matter
particles of $1.17\times10^7\msun$ each and $4\times10^9$ gas cells.
It uses the Eulerian AMR code RAMSES \citep{teyssier02}, which is based
on a graded octree structure with a cell by cell refinement. 
\adb{Each cell is refined if the dark matter mass exceed 8 times the dark 
matter particle mass or if the gas mass exceeds 8 times the initial gas mass 
resolution.  Shocks and contact discontinuities are not refined.
The former has been shown \citep{teyssier02} not to lead to spurious effects
in the cosmological context. The latter could lead to an underestimate the 
subsonic turbulence induced by gravitational collapse, but it has been
estimated to stay below 10-15\% of the total thermal or gravitational energy
\citep[][and references therein]{vazza11}.}
The refinement
criteria are based on gradients of the flow variables in a given cell. 
The N-body solver uses a Particle-Mesh scheme and the Poisson equation is 
solved using a multigrid solver. The hydrodynamical solver uses an unsplit 
second order Godunov method.

The cosmological model used is the $\Lambda$CDM model with $\oml=0.7$,
$\omm=0.3$, $\omb=0.045$, $h=0.7$ and $\sigma_8=0.95$. Dark matter haloes are
identified using the AdaptaHop algorithm. 
The physical processes for galaxy formation implemented in the RAMSES code
include radiative cooling, UV background radiation, star formation and SN
feedback. 
Gas can cool radiatively to a minimum temperature of $10^4 K$ with a rate
depending on the local metallicity. Star formation is included for gas above
a threshold density $n_{H}>n_0$, with $n_0=0.1 \cmc$, designed to match the
Kennicutt SFR law with an efficiency of $5\%$ per free-fall time. 
UV heating is included using the Haardt and Madau background model. 
SN feedback and the associated metal enrichment are implemented based on the 
blast-wave model described in \citep{dubois08}, where 50\% of the SN energy 
is deposited as bulk motion in a gas bubble of a certain radius,
and the other half is assumed to be radiated away.
High-density regions are described by a polytropic equation of state with
$\gamma=5/3$ to model the complex, multi-phase structure of the ISM.



\section{Maximum influx plane}
\label{sec:strip}

In \se{fit} we described our main algorithm for defining the stream plane at
a given spherical shell as the best fit great circle to the angular positions 
of the three dominant streams.  
An alternative way to defining the plane is by maximizing the influx through 
a belt about the great circle \citep[e.g.,][]{aubert04}. 
This includes the influx in all streams
as well as the pancakes.
In a concrete example, we use a width of $\pm \pi/9$ for this ring, 
such that it covers about one third of the spherical shell. 
This algorithm has been applied to the whole sample of haloes, in 
a spherical shell of thickness $0.1\Rv$ about $r=\Rv$. 

\Fig{strip} shows the distribution of the cosine of the angle between this 
maximum-influx plane and the stream plane as defined in our main analysis, 
\se{fit}. 
We find that in the vast majority of the haloes the stream planes defined by
the two methods practically coincide, with a median of $\cos\theta = 0.96$
for the angle between the two planes.
This is an encouraging evidence for the robustness of our analysis.

\section{Simulation axes bias}
\label{sec:axes}

A potential caveat in our analysis of preferred planes is the numerical 
tendency for artificial alignment of these planes with the simulation grid.
A multigrid Poisson solver, and in particular the hydrodynamical solver,
create non-physical forces along the preferred Cartesian directions of the
simulation grid, which act to align the mass distribution with the grid 
\citep{hahn10}.
This is especially relevant for the galactic disc plane, which involves
scales not much larger than the resolution scale, but it may also propagate
to the sheets on larger scales.
This artificial alignment is expected to be stronger at lower redshifts, where
the discs might have had enough time to relax to the closest grid direction,
so we expect our analysis of massive galaxies at $z=2.5$ to be less
vulnerable to this numerical effect, despite the 1-kpc resolution.

In \fig{sim_bias_aitoff}, for each of three planes in question,
we display the distribution of directions of the normals to this plane 
(a blue dot per galaxy).
The planes are 
the disc plane, the stream plane and the angular momentum at $\Rv$.
The positions of the six simulation grid axes are marked (red dots).
A tendency for numerical alignment would appear as clustering of the
blue points about the red points.
For the three planes in question, the distribution of blue points
appears to be isotropic.
This suggests that the numerical alignment is weak, even for the disc plane.

In \fig{sim_bias_pdf} we show the PDF of the cosine of the minimum
angle between the normal to the plane in question
and the any of the simulation grid axes.
The count in each bin of $\cos \theta$ is associated with a Poisson error bar.
This distribution is compared to a null hypothesis of isotropic distribution
of plane normals.
For the disc plane, a KS-test marginally rejects the null hypothesis with 
a p-value of 0.02. Inspecting the PDF, we see a significant deficiency of 
counts in the range $\cos\theta =0.80-0.88$ involving $\sim 10\%$
of the galaxies, and an excess involving
a similar fraction of the galaxies at $\cos\theta =0.96-1.00$.
We interpret this as an offset of $\Delta \cos\theta \sim 0.08$ involving
$\sim 20\%$ of the galaxies.
The overall shift of the median $\cos\theta$ compared to the random
distribution of plane normals is about $0.01$.
This small effect represents the level of error that we should assign to
any measure of alignment between the disc and other planes.
The stream plane and the angular momentum at $\Rv$ do not show a noticeable
numerical alignment with the simulation grid. The KS-test p-values are 0.83
and 0.5 respectively, and the shifts in the median $\cos\theta$ are less 
than $0.01$.


\section{Poor stream planes}
\label{sec:bad}

\Fig{bad_planarity} shows the 3 worst cases for a stream plane. 
The streams in the middle and right panels do not lie on one plane that
includes the halo centre. On the other hand, the halo shown in the left panel
has three main streams that do define a plane and are embedded in a visible
pancake, but it also shows two other streams that do not lie on the same plane.

\label{lastpage}

\end{document}